\documentclass[acmsmall]{acmart}
\AtBeginDocument{%
  }

\setcopyright{acmlicensed}
\copyrightyear{2025}
\acmYear{2025}
\acmDOI{XXXXXXX.XXXXXXX}
\acmConference[Conference acronym 'XX]{Make sure to enter the correct
  conference title from your rights confirmation email}{June 03--05,
  2025}{Woodstock, NY}
\acmISBN{978-1-4503-XXXX-X/2018/06}

\usepackage{ulem}
\usepackage{tabularx}
\usepackage[table]{xcolor}
\usepackage{multirow}
\usepackage{pifont}
\usepackage{manfnt}
\usepackage{tikz}
\usepackage{longtable}
\usepackage{needspace}

\usepackage{mdframed}

\newmdenv[
  backgroundcolor=gray!10,
  linecolor=gray,
  linewidth=1pt,
  roundcorner=5pt
]{definitionbox}

\newcommand{\nff}[1]{\textcolor{black}{#1}}
\newcommand{\papers}{140}

\begin{document}

\title{Feature Request Analysis and Processing: Tasks, Techniques, and Trends}

\author{Feifei Niu}
\email{feifeiniu96@gmail.com}
\orcid{0000-0002-4123-4554}
\author{Chuanyi Li}
\email{lcy@nju.edu.cn}
\orcid{0000-0001-9270-5072}
\authornote{Corresponding Author.}
\author{Haosheng Zuo}
\email{211250074@smail.nju.edu.cn}

\author{Jionghan Wu}
\email{652024320005@smail.nju.edu.cn}


\affiliation{
  \institution{State Key Laboratory for Novel Software Technology, Nanjing University}
  \city{Nanjing}
  \state{Jiangsu}
  \country{China}
}
\author{Xin Xia}
\email{xin.xia@acm.org}
\orcid{0000-0002-6302-3256}
\affiliation{
   \institution{The State Key Laboratory of Blockchain and Data Security, Zhejiang University}
   \city{Hangzhou}
   \state{Zhejiang}
   \country{China}
}

\renewcommand{\shortauthors}{Niu et al.}

\begin{abstract}
Feature requests are proposed by users to request new features or enhancements of existing features of software products, which represent users' wishes and demands. Satisfying users' demands can benefit the product from both competitiveness and user satisfaction. Feature requests have seen a rise in interest in the past few years and the amount of research has been growing. However, the diversity in the research topics suggests the need for their collective analysis to identify the challenges and opportunities so as to promote new advances in the future. In this work, following a defined process and a search protocol, we provide a systematic overview of the research area by searching and categorizing relevant studies. We select and analyze \papers~primary studies using descriptive statistics and qualitative analysis methods. We classify the studies into different topics and group them from the perspective of requirements engineering activities. We investigate open tools as well as datasets for future research. In addition, we identify several key challenges and opportunities, such as: (1) ensuring the quality of feature requests, (2) improving their specification and validation, and (3) developing high-quality benchmarks for large language model-driven tasks. \nff{Our artifacts are publicly available at \url{https://github.com/feifeiniu-se/FeatureRequestSurvey}.}
\end{abstract}


\setcopyright{cc}
\setcctype{by}
\acmJournal{TOSEM}
\acmYear{2026} \acmVolume{1} \acmNumber{1} \acmArticle{}
\acmMonth{1} \acmDOI{10.1145/3799899}


\received{31 July 2025}
\received[revised]{20 December 2025}
\received[accepted]{11 Feb 2026}

\maketitle

\section{Introduction}
Software feature is defined as a prominent or distinctive user-visible aspect, quality, or characteristic of a software system or systems~\cite{kang1990feature}. A feature can be briefly defined as a functionality (or behavior) that a software product should provide~\cite{batory2005feature, benavides2010automated}. A feature request is, in turn, a request filed by the user requesting for new feature or enhancement to software functions.

Feature requests are usually proposed by users, conveying users requirements. There is a lot of valuable information contained in users' requests for software maintenance. According to Iacob and Harrison~\cite{iacob2013retrieving}, approximately 23.3\% of app reviews are feature requests, among which some are users' suggestions for new features, while others indicate users' preferences for redesigning existing app features. Recent research gradually considers users' requests as a type of software requirements~\cite{rath2018analyzing, cai2025automatic}. They can facilitate communication between developers, testers, and users, and also serve as a coordination mechanism for multiple stakeholders and the software team~\cite{bertram2009social, bertram2010communication}. Thus, implementing user requests can help improve users' satisfaction, which is a pivotal factor in reducing the possibility of software failure, according to the Standish report~\cite{clancy1995standish}. What's more, implementing feature requests in an early stage allows the product to provide more satisfying features to the users, and can effectively attract more users to the product, which makes the product stand out among plenty of similar products~\cite{dalpiaz2019re}.

Due to the considerable value of feature requests to software, the analysis and processing of feature requests in software has been widely studied in practice. Recent years have witnessed a surge research interest in feature requests, such as identifying feature requests from issues reports~\citep{iacob2013retrieving, laura2013analysis, roger2017preliminary, venkatesh2018app}, extracting software features from users' requests~\citep{bakar2016extracting, elsa2017which, timo2017safe, licorish2017attributes, dalpiaz2019re}, feature requests prioritization~\citep{kegel2023automating, de2024opinion, salleh2024review}, and locating features in the source code~\citep{palomba2017recommending, panichella2016app, ciurumelea2017analyzing, kato2017user}. These have greatly promoted the application of feature requests in software engineering. Furthermore, to investigate the challenges and opportunities in software change request repositories, Cavalcanti et al.~\cite{cavalcanti2014challenges} conducted a systematic mapping study based on two dimensions: challenges and opportunities related to change request repositories. Tavakoli et al.~\cite{tavakoli2018extracting} surveyed the state-of-the-art (SOTA) tools and techniques employed in detecting informative aspects in mobile application reviews. Bakar et al.~\cite{bakar2015feature} provided a systematic literature review of the SOTA approaches in feature extractions from natural language requirements for reuse in software product line engineering. Wang et al.~\cite{wang2019systematic} investigated the implicit feedback and explicit feedback employed for crowdsourced requirements engineering (RE) activities. Most existing secondary studies have focused on narrow aspects of feature requests, such as analyzing app store reviews for feature requests or addressing specific tasks like feature extraction. However, to the best of our knowledge, there is no comprehensive study that provides a holistic view of the entire research landscape on feature requests across different tasks and contexts. Furthermore, the field lacks an integrated discussion of overarching challenges and open research opportunities that span the full lifecycle of feature request management. Additionally, there is a lack of understanding regarding the frequency and location of studies published over the past few years.

\begin{figure}[t!]
	\centering
		\includegraphics[width=0.9\linewidth]{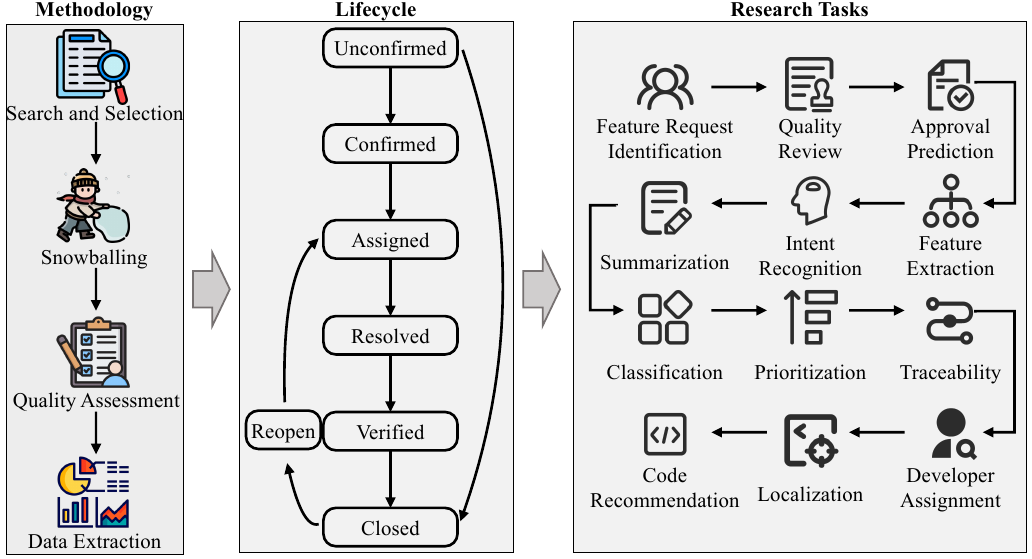}
	 \caption{Study Overview.}\label{fig:overview}
\end{figure}

In this paper, we conduct a systematic literature review on feature requests to portray the landscape of research practices from a requirements engineering perspective and pinpoint research challenges and opportunities, following the guidelines of~\cite{petersen2008systematic, petersen2015guidelines}. We propose four research questions and define a set of search strings to retrieve the initial relevant literature. Then snowballing is employed to retrieve the whole literature. Finally, a total of \papers~literature studies are selected. In particular, the studies are scrutinized for feature requests analysis and processing for the software maintenance. We report on the demographics of the published articles to provide a comprehensive view. 
Furthermore, we also categorize the literature according to the research topic and provide a detailed analysis of each topic. Figure~\ref{fig:overview} presents an overview of the study topics in relation to requirements activities, encompassing 12 distinct research topics. We also expand each research topic with an explanation of the research progress, accompanied by a demonstration of the proposed approach. To facilitate future studies, we investigate the public tools and datasets for feature requests. Finally, we pinpoint the research gaps, which can help researchers understand the road ahead in investigating user requests. 

The remainder of this paper is organized as follows: Section~\ref{relatedwork} compares this study with related work. Section~\ref{methodology} presents the research methodology and details the review process. In Section~\ref{sec:demographics}, we outline the demographics of the reviewed literature. Section~\ref{sec:analysis} provides a detailed review of the literature on feature request analysis and processing, encompassing 12 distinct research topics. Section~\ref{subsec:datasetandtool} summarizes existing tools and open datasets on feature requests.
Section~\ref{sec:challenges} discusses the challenges and opportunities for future research. Section~\ref{threats} addresses threats to validity. Finally, Section~\ref{conclusion} concludes the paper and outlines directions for future work.

\section{Preliminary and Related Work}\label{relatedwork}

\subsection{Feature Request}

\textit{A feature request is a submission from a customer, stakeholder, or employee proposing the addition of a specific capability to a product or service.~\footnote{\url{https://www.atlassian.com/agile/product-management/feature-request}}}

\begin{figure}
    \centering
    \includegraphics[width=0.9\linewidth]{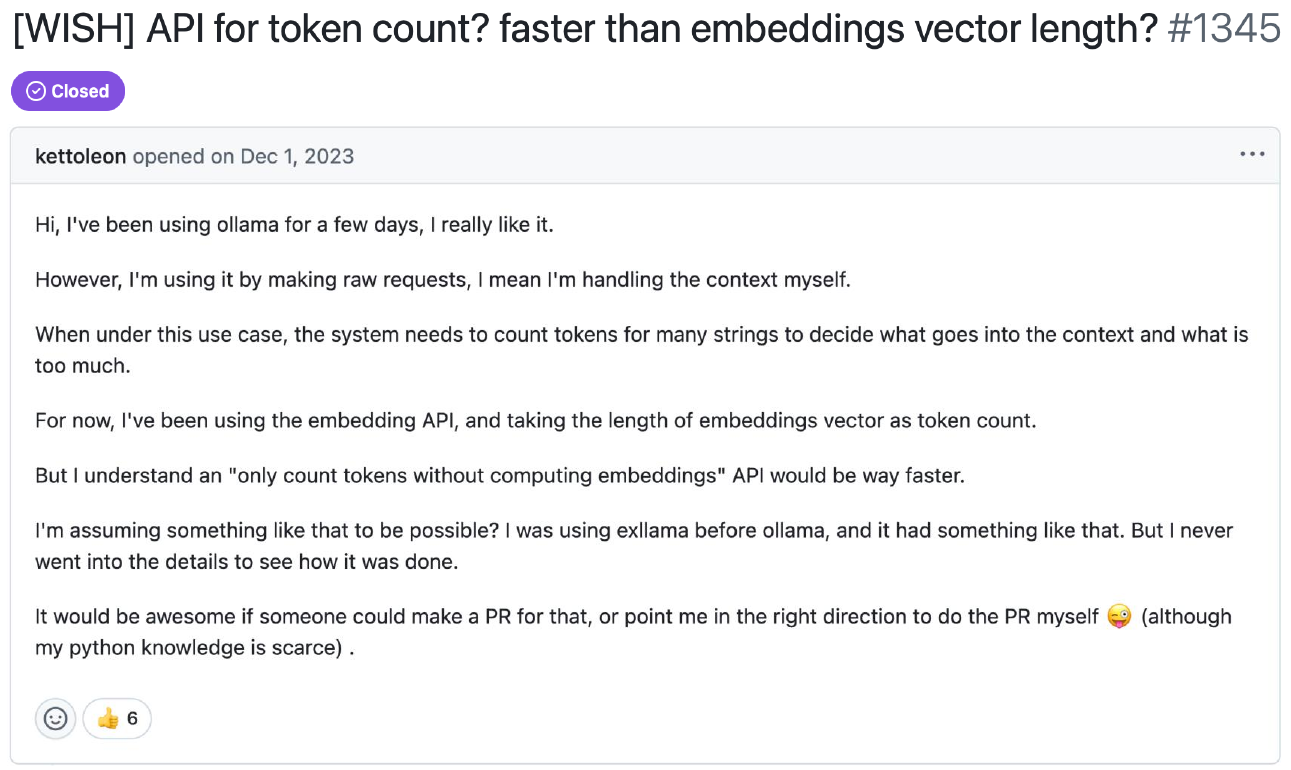}
   \caption{An example feature request from the Ollama project (\href{https://github.com/ollama/ollama/issues/1345}{Issue \#1345})}
    \label{fig:example}
\end{figure}

Figure~\ref{fig:example} illustrates a real-world feature request (\href{https://github.com/ollama/ollama/issues/1345}{Issue~\#1345}) from the Ollama project~\footnote{\url{https://github.com/ollama/ollama/}}, where a stakeholder proposed an API for counting tokens in the system. This request was ultimately labeled as \textit{Closed}. In such submissions, authors often describe the context motivating the feature, detail the proposed functionality, and outline its potential benefits. Feature requests constitute one of the most important sources of software requirements~\cite{cai2025automatic}, particularly in user-centered and fast-paced development contexts. 

As depicted in Figure~\ref{fig:overview}, the lifecycle of a feature request closely parallels that of a bug report~\cite{d2007bug}, typically progressing through the stages of \textit{unconfirmed}, \textit{confirmed}, \textit{assigned} to developers, \textit{resolved}, and finally \textit{verified} and \textit{closed}. In some cases, a request may be \textit{reopened} and \textit{reassigned}. While a bug report typically describes an unexpected or erroneous system behavior, a feature request instead proposes new or enhanced functionality that does not currently exist in the system, aiming to extend its capabilities or improve the user experience~\cite{nyamawe2019automated}. Across this lifecycle, numerous research topics have been explored, including feature request identification, classification, and prioritization.

\subsection{Related Work}
Before describing our work, we provide an overview of related systematic reviews and mapping studies. To the best of our knowledge, there is no existing literature on comprehensive feature request analysis and processing. Although there are numerous reviews addressing user feedback, they seldom delve deeply into feature requests. Therefore, we conducted a systematic review study to address this gap. 

Cavalcanti et al.~\cite{cavalcanti2014challenges} performed a systematic mapping study in 2014 to characterize the research challenges and opportunities for software change request repositories. The result was a two-taxonomy based on two dimensions: challenges and opportunities related to change request repositories. However, the search strings of this paper mainly involve ``change request'' and ``bug report'', etc., not ``feature request''. Thus, some feature request-related research may be missed. Tavakoli et al.~\cite{tavakoli2018extracting} surveyed the SOTA tools and techniques used to detect valuable information concerning user needs and requirements from mobile application reviews. User reviews of mobile applications may contain bug reports, user experience, and user requirements, which are pivotal information relevant to app development and improvement. Therefore, they surveyed the mining tools that have been used to unleash the value of app reviews for mobile app development. They mainly focus on obtaining useful information from user reviews, which our study pays attention to the analysis and processing of user feature requests, which is downstream of their survey. Wang et al.~\cite{wang2019systematic} provided a mapping study of literature on crowdsourced user feedback employed for requirements engineering purposes. They provided an overview of the types of user feedback that have been employed for crowdsourced requirements engineering activities. They investigated both implicit feedback and explicit feedback. However, they only mapped the literature to requirements engineering activities without further investigation into the SOTA techniques for each topic. Bakar et al.~\cite{bakar2015feature} provided a systematic literature review of the SOTA approaches in feature extractions from natural language requirements for reuse in software product line engineering. \nff{D{\k{a}}browski et al.~\cite{dkabrowski2022analysing} conducted a systematic review of 182 studies on app review analysis for software engineering. They categorized existing work into feature request identification, bug report detection, sentiment analysis, and user experience evaluation, and summarized commonly used techniques such as machine learning, deep learning, and topic modeling. The study highlighted how app review analysis supports requirements, design, testing, and maintenance. However, their work remains a high-level categorization, lacking an in-depth methodological synthesis for each task. In contrast, our study discusses feature request analysis from the perspective of the software development lifecycle, providing a more comprehensive understanding of how feature requests are generated, processed, and utilized throughout different development stages.} \nff{Dit et al.~\cite{dit2013feature} conducted a comprehensive survey on feature location in source code, providing a detailed taxonomy that categorizes 89 studies according to multiple analytical dimensions, including static, dynamic, textual, historical, and hybrid approaches.
The survey summarizes key techniques for locating the implementation of software features and identifies major research gaps, such as the lack of standardized benchmarks, inconsistent evaluation practices, and limited cross-language applicability.}
Bischoff et al.~\cite{bischoff2019integration} created a panoramic view of the current literature to pinpoint gaps and supply insights in the integration of feature models. They sought to understand, characterize, and summarize the literature about the integration of feature models. Recently, Cai et al.~\citep{cai2025automatic} reviewed existing work on requirements elicitation from user-generated content, categorized studies by data sources, methods, and representations, and proposed a research framework to guide future work in this area.


\nff{Although several secondary studies have investigated concepts related to user feedback and software features, most have focused on adjacent topics or partial perspectives rather than offering a systematic synthesis of feature request research across the software development lifecycle. Existing work has examined areas such as change request repositories~\cite{cavalcanti2014challenges}, requirement or information extraction from app reviews~\cite{tavakoli2018extracting, cai2025automatic}, feature modeling~\cite{bischoff2019integration}, feature extraction in software product lines~\cite{bakar2015feature}, feature location~\cite{dit2013feature}, and high-level mapping studies on crowdsourced feedback~\cite{wang2019systematic, dkabrowski2022analysing}. However, these studies do not provide a unified view of how feature requests are managed throughout the software lifecycle.}

\nff{In contrast, our study presents a lifecycle-oriented synthesis of feature request research. We categorize existing work according to key stages of the software development process and, within each stage, analyze the techniques and datasets employed. This perspective delivers a comprehensive and methodical understanding of how feature requests emerge, evolve, and contribute to continuous software evolution, offering a cohesive view that bridges fragmented prior research.}

\section{Research Methodology} \label{methodology}
In this paper, we conduct a systematic literature review following the guidelines proposed by Petersen et al.~\cite{petersen2008systematic, petersen2015guidelines}. The review process consists of three main phases: planning, execution, and analysis. During the planning phase, we defined the research questions and established the survey protocols aligned with our study’s objectives. In the execution phase, we applied these protocols to select relevant studies, employing snowballing techniques to ensure comprehensive coverage of literature related to feature requests. Finally, in the analysis phase, four co-authors collaboratively reviewed the selected papers to address the research questions. The overall research process is illustrated in Figure~\ref{fig:overview}.

\subsection{Research Questions}

\begin{itemize}
    \item \textbf{RQ1: What are the demographics of the published articles?} 
    
    \textbf{Goal} — This research question aims to identify \textbf{when} the articles were published and how the number and frequency of studies have evolved; \textbf{where} feature request analysis and processing research has been published; \textbf{what} types of articles (academic or industry) exist; and \textbf{how} these studies address challenges related to feature requests.
    
    \item \textbf{RQ2: How do existing studies analyze and process feature requests?}
    
    \textbf{Goal} — This question investigates current \textbf{research practices} regarding feature requests, providing a comprehensive overview from a requirements engineering perspective.
    
    \item \textbf{RQ3: What publicly available datasets and tools are presented in the literature?}
    
    \textbf{Goal} — This question seeks to identify open-source \textbf{datasets and tools}, categorize them according to research topics, and facilitate their future use.
    
    \item \textbf{RQ4: What are the research challenges and opportunities related to feature requests?}
    
    \textbf{Goal} — This question explores the existing \textbf{challenges} in feature request analysis and management, while highlighting potential avenues for future research.
\end{itemize}




\subsection{Search and Selection Processes}

\subsubsection{Search Strategy}
\begin{figure}
    \centering
    \includegraphics[width=\linewidth]{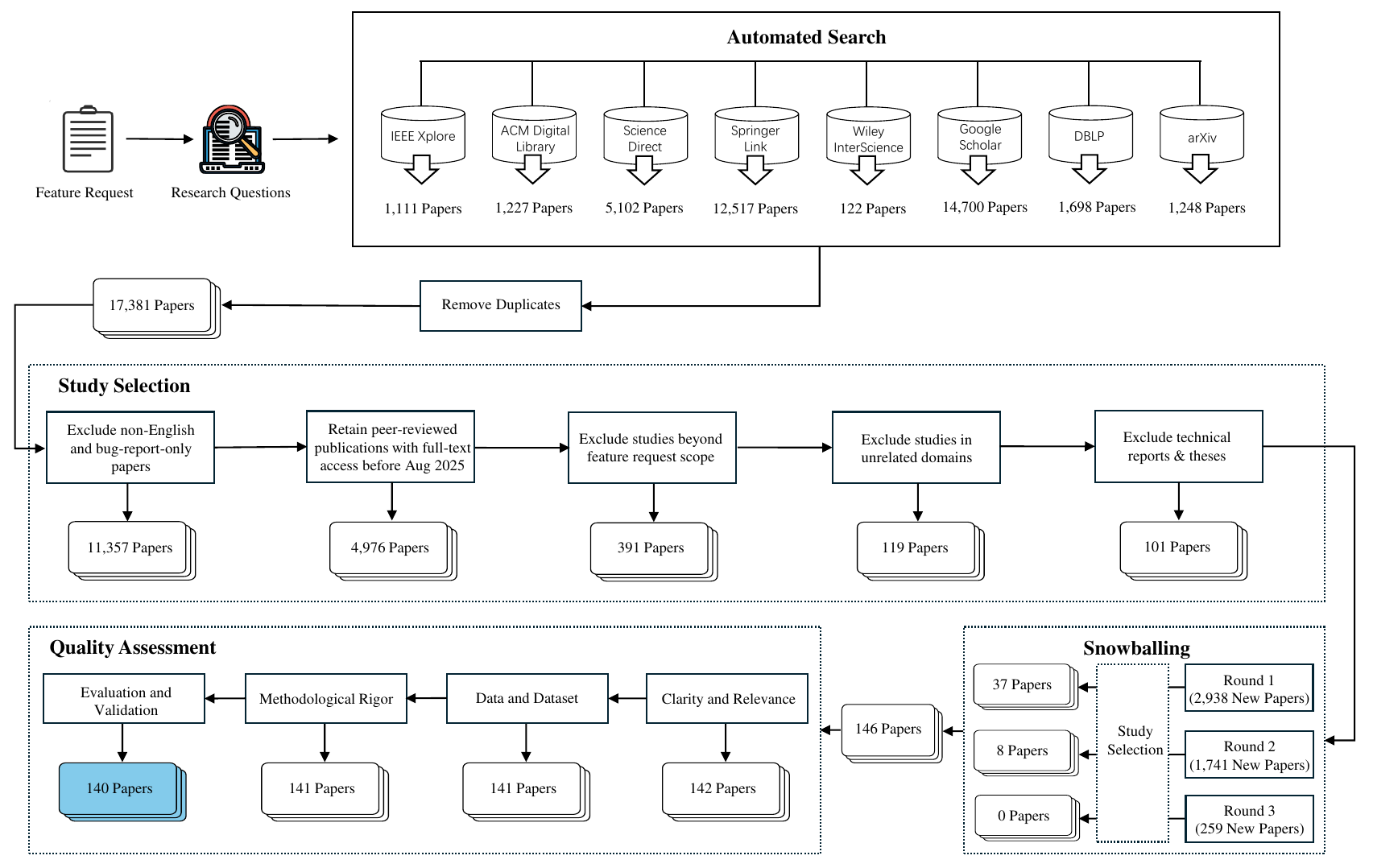}
   \caption{Study Search and Selection Process.}
    \label{fig:searchprocess}
\end{figure}

\nff{As illustrated in Figure~\ref{fig:searchprocess}, we adopted a systematic search strategy combining both automated searching and snowballing to ensure comprehensive coverage of related studies while minimizing the risk of missing relevant work. To address our research questions, we first identified the following keywords related to the feature request topic:}

\textit{Keywords related to feature request}: \textit{feature request}, \textit{user request}, \textit{user requirement}, \textit{enhancement report}, \textit{enhancement request}, \textit{just-in-time requirement}

\nff{The same search keywords were applied across all digital libraries, with only minor syntactic adaptations to comply with each database’s specific query requirements. Two authors independently conducted the automated search, which helps avoid bias and ensures reliability. The search was performed across the digital libraries listed in Figure~\ref{fig:searchprocess}, covering all available publication years up to August 2025. After merging and removing duplicates, 17,381 distinct papers were obtained.}

\subsubsection{Inclusion/Exclusion Criteria}
To identify the most relevant studies on feature request analysis and processing, we drew upon prior work~\citep{giray2023use, zakari2020multiple, croft2022data, niu2025deep, hu2025assessing} and established a set of inclusion (ICs) and exclusion criteria (ECs), summarized in Table~\ref{tab:criteria}. These criteria were applied to the titles, abstracts, and keywords of candidate papers to ensure relevance and quality.

Papers not written in English (IC1) were excluded at the outset, as were those focusing solely on bug reports (EC2), which lie outside the scope of our feature request study. We then ensure that the remaining papers were published in peer-reviewed venues --- such as journals, conferences, or workshops --- before August 2025 (IC2) with full-text access available (IC3). Studies that did not address feature requests or enhancement reports within the context of software requirements were also excluded (IC4). Additionally, papers centered on user requests in unrelated domains, such as marketing or recommendation systems, were excluded (EC1) due to misalignment with our focus. To maintain the rigor and consistency of the review, technical reports and theses (EC3) were omitted, as they often lack formal peer review. Given the field’s emerging nature and the ongoing publication process of many works, we included recent preprints from arXiv to capture the latest developments.

Through this rigorous screening process, we refined a comprehensive yet focused set of relevant studies for detailed analysis. Overall, our study pool consists of 101 selected papers. 

\begin{table}[t!]
\centering
\small
\caption{Inclusion and Exclusion Criteria.}\label{tab:criteria}
\begin{tabular}{m{0.1\linewidth}<{\raggedright}m{0.8\linewidth}<{\raggedright}}
\toprule
\multicolumn{2}{l}{Inclusion Criteria}                                                                                                                  \\ \midrule
\rowcolor[HTML]{EFEFEF} 
IC1       & The paper is written in English.                                                                                 \\
IC2       & The paper is published before August 2025.                                                                     \\
\rowcolor[HTML]{EFEFEF} 
IC3       & The paper is published in a peer-reviewed journal, conference proceedings, or a workshop, with the full-text available, including recent publications available on arXiv.            \\
IC4       & The paper addresses issues related to feature requests/enhancement reports in RE activities.                    \\ \midrule
\multicolumn{2}{l}{Exclusion Criteria}                                                                                                                  \\ \midrule
\rowcolor[HTML]{EFEFEF} 
EC1 & The paper is about the application of user requests for non-RE activities,   like marketing, recommending apps, etc. \\
EC2 & The paper primarily addresses issues related to bug reports.\\
\rowcolor[HTML]{EFEFEF}
EC3 & The paper is a technical report or thesis.\\
\bottomrule
\end{tabular}
\end{table}






\subsubsection{Snowballing}
To ensure a comprehensive literature collection, we performed both forward and backward snowballing into our study selection process, following the guidelines outlined by Wohlin~\cite{wohlin2014guidelines}. This iterative process involved examining the references and citations of the selected studies in successive rounds. 
\nff{After completing the database search and study selection (resulting in 101 papers), we performed iterative forward and backward snowballing on the selected studies. As shown in Figure~\ref{fig:searchprocess}, a total of three rounds of snowballing were conducted, which retrieved 2,938, 1,741, and 259 papers in each round, respectively. These papers were screened using the same inclusion and exclusion criteria presented in Table~\ref{tab:criteria}, resulting in 37, 8, and 0 newly included papers in each iteration. Ultimately, 45 additional primary studies were retained, increasing the final number of included studies from 101 to 146. This process ensured that relevant studies that may have been missed during the automated search were thoroughly captured.}

\subsubsection{Quality Assessment}

Ensuring the quality of included studies is essential to provide a reliable overview of feature request research. To achieve this, we developed a quality assessment checklist based on three main dimensions: the quality of the data, the robustness of the modeling techniques, and the rigor of the evaluation methods, as summarized in Table~\ref{tab:quality-checklist}.


\nff{
Each criterion was assessed using a binary scoring scheme (satisfied = 1, not satisfied or unclear = 0). Four authors independently evaluated all 146 candidate studies. In cases of disagreement, the reviewers discussed the assessment results until consensus was reached.
For each study, we calculated an overall quality score by summing the scores across all eight criteria, resulting in a maximum possible score of 8. Following common practices in secondary studies, we excluded papers that satisfied fewer than 70\% of the quality criteria (i.e., a total score lower than 6). These studies typically lacked essential information on data sources, methodological details, or evaluation procedures, which would hinder reliable data extraction and synthesis.
As a result of this quality assessment process, six papers were excluded for not meeting the established quality threshold, leaving \papers~high-quality studies for detailed data extraction and further analysis. For transparency, the list of excluded papers and the corresponding unmet criteria are reported in our GitHub repository.
}

\begin{table}[t!]
\centering
\small
\caption{Quality Assessment Checklist for Feature Request Studies.}
\label{tab:quality-checklist}
\begin{tabular}{p{3.5cm}p{1cm}p{8.5cm}}
\toprule
\textbf{Category}                    & \textbf{ID} & \textbf{Criteria} \\ 
\midrule
\multirow{4}{*}{Clarity and Relevance}         
& CR1  & Does the paper clearly describe its objectives, scope, and relevance to feature requests within requirements engineering? \\[4pt] \cline{3-3}
& CR2  & Are the research questions clearly defined and aligned with the study’s goals?  \\[6pt] \midrule

\multirow{4}{*}{Data and Dataset}           
& DD1  & Does the study explicitly report the source and nature of feature request data? \\[4pt] \cline{3-3}
& DD2  & Are details of the dataset, including size, granularity, and data collection methods, clearly reported? \\[6pt] \midrule

\multirow{4}{*}{Methodological Rigor}   
& MR1  & Is the approach or methodology for analyzing feature requests clearly described and scientifically rigorous? \\[4pt] \cline{3-3}
& MR2  & Are feature extraction, classification, or prioritization methods clearly and comprehensively demonstrated? \\[6pt] \midrule

\multirow{4}{*}{Evaluation and Validation}  
& EV1  & Does the paper clearly explain how the evaluation or validation was conducted? \\[4pt] \cline{3-3}
& EV2  & Are performance metrics or validation criteria clearly reported and justified? \\[6pt]
\bottomrule
\end{tabular}
\end{table}

\subsection{Data Extraction and Synthesis}

\noindent
\paragraph{Data Extraction}

After determining the final study pool, we proceeded to the data extraction phase, systematically gathering information from the selected studies, aligning closely with our defined research questions (RQs). To maintain consistency and comprehensiveness, we developed a structured data extraction form explicitly linked to each RQ, as detailed below:

\begin{itemize}
    \item \textbf{Demographic Information (RQ1):}
        \begin{itemize}
            \item Author(s), year of publication, paper title, and publication venue (e.g., journal, conference, workshop, arXiv).
            \item Type of study (academic research or industry report).
            \item Main challenges or problems addressed by the study in the context of feature requests.
        \end{itemize}

    \item \textbf{Analysis and Processing Practices (RQ2):}
        \begin{itemize}
            \item Specific methods and approaches employed for analyzing, classifying, prioritizing, summarizing, or managing feature requests.
            \item Data preprocessing techniques and vectorization methods are described.
            \item Evaluation criteria and performance metrics used to validate proposed methods.
        \end{itemize}

\item \textbf{Publicly Available Datasets and Tools (RQ3):}
\begin{itemize}
    \item Availability and accessibility of datasets (including sources, size, granularity, and usage contexts).
    \item Open-source tools, scripts, or platforms provided or utilized by the study, including URLs or repository links.
    \item Categorization of the datasets and tools according to their purposes or research themes.
\end{itemize}

\item \textbf{Challenges and Opportunities (RQ4):}
\begin{itemize}
    \item Reported limitations, difficulties, or gaps encountered by researchers when analyzing or managing feature requests.
    \item Potential areas for future improvement or research directions explicitly mentioned by the authors.
    \item Any recommendations or insights provided for overcoming identified challenges.
\end{itemize}
\end{itemize}

Four co-authors independently performed the data extraction process, with any discrepancies or conflicts resolved through discussion and consensus to ensure the accuracy and reliability of the extracted information.

\noindent

\nff{\paragraph{Data Synthesis.}
The goal of this survey is to aggregate evidence from primary studies and provide an overview of the state of the art on feature request analysis and processing. Following the systematic review guidelines~\cite{keele2007guidelines}, we synthesized the extracted data using a combination of descriptive (quantitative) synthesis and qualitative thematic synthesis, aligned with our four research questions.
For \textbf{RQ1 (Demographics)}, we performed descriptive synthesis on bibliographic and contextual attributes (e.g., publication year, venue type, and study type). We report the evolution of the literature over time and the distribution across venues using frequency counts. 
For \textbf{RQ2 (Analysis and Processing Practices)}, we synthesized the extracted methodological information by grouping studies according to the main feature request tasks they target (e.g., classification, prioritization, summarization, management) and the approaches they employ. We summarize the prevalence of different practices (including preprocessing, vectorization/representation, and evaluation setups) using structured comparison tables, and we narratively discuss commonalities and differences across studies.
For \textbf{RQ3 (Public Datasets and Tools)}, we first summarized availability-related information (e.g., whether datasets/tools are publicly accessible, their sources) in a structured catalog. We then conducted a qualitative categorization of the datasets and tools by their intended purposes and research themes, and reported results for each category to facilitate reuse.
For \textbf{RQ4 (Challenges and Opportunities)}, we conducted an inductive thematic synthesis. During data extraction, any text explicitly describing a limitation, difficulty, research gap, or future work was copied verbatim into the extraction form. Two reviewers independently performed open coding on these excerpts to identify recurring concepts and iteratively refined a shared codebook through discussion. Conceptually similar codes were merged into higher-level themes, which were manually reviewed to ensure consistent granularity and clear boundaries. We report the resulting themes together with representative evidence from primary studies and, where appropriate, the number of studies mentioning each theme. When divergent or conflicting observations emerged, we explicitly described them rather than forcing a single conclusion.
}

\section{Demographics} \label{sec:demographics}
\begin{figure}[t!]
  \centering
  \includegraphics[width=0.9\textwidth]{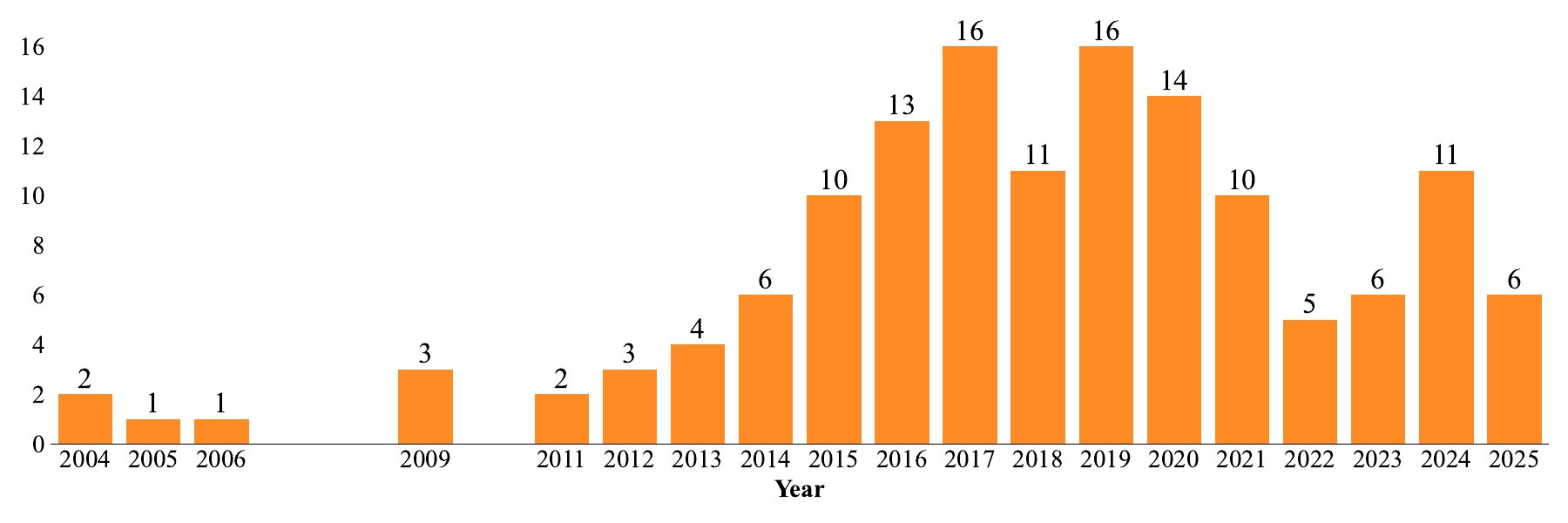}
  \caption{Number of Primary Studies over the Years.}
  \label{fig:year}
\end{figure}

\begin{figure}[t!]
  \centering
  \includegraphics[width=0.9\textwidth]{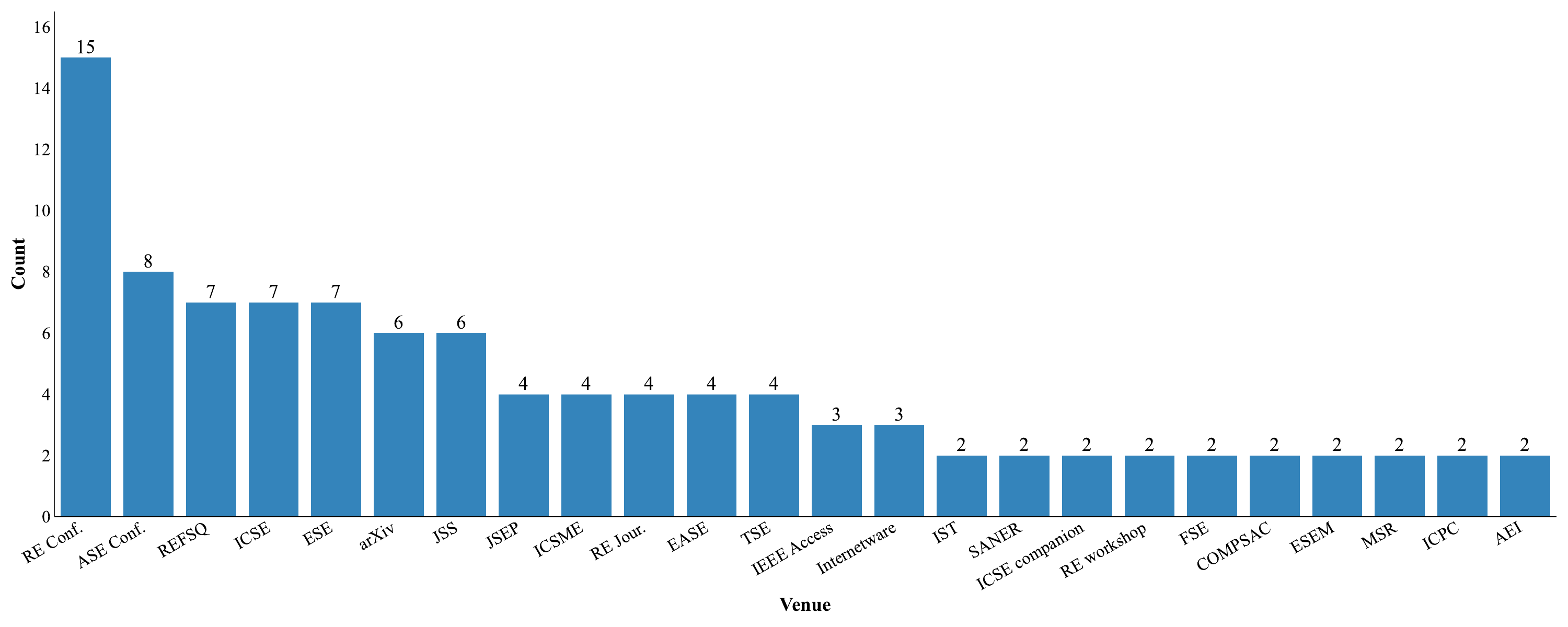}
  \caption{Top Publication Venues.}
  \label{fig:venue}
\end{figure}

A total of \papers~research papers were included in the final study set, covering publications up to August 2025. As shown in Figure~\ref{fig:year}, the number of primary studies has increased significantly over time, showing a clear upward trend from 2011 to 2017, with the peak occurring in 2017 and 2019 (16 studies). Activity remained relatively high through 2020, followed by a decline likely due to shifting research priorities. A notable resurgence occurred in 2024 (11 studies), coinciding with the rise of LLMs, which introduced new research opportunities. Overall, the data reveals cyclical research activity with notable peaks and dips over the years. Figure~\ref{fig:venue} presents the distribution of the selected papers across journals and conferences, highlighting the most frequently appearing publication venues. A total of 88 papers were published in conferences, 46 in journals, and 6 as preprints. Most studies were published in top-tier software engineering outlets, particularly in the field of RE. High-frequency venues include International Conference on Requirements Engineering (RE Conf., 15 papers), International Conference on Automated Software Engineering (ASE Conf., 8 papers), International Working Conference on Requirement Engineering: Foundation for Software Quality (REFSQ, 7papers), International Conference on Software Engineering (ICSE, 7 papers), Empirical Software Engineering (ESE, 7 papers), Journal of Systems and Software (JSS, 6 papers), Transactions on Software Engineering (TSE, 4 papers),
International Conference on Software Maintenance and Evolution (ICSME, 4 papers), Requirements Engineering Journal (RE Jour., 4 papers), and International Conference on Evaluation and Assessment in Software Engineering (EASE, 4 papers). These results confirm that the topic has received strong attention from the software engineering community.

\begin{figure}[t!]
  \centering
  \begin{minipage}[t]{0.5\textwidth}
    \centering
    \includegraphics[width=\textwidth]{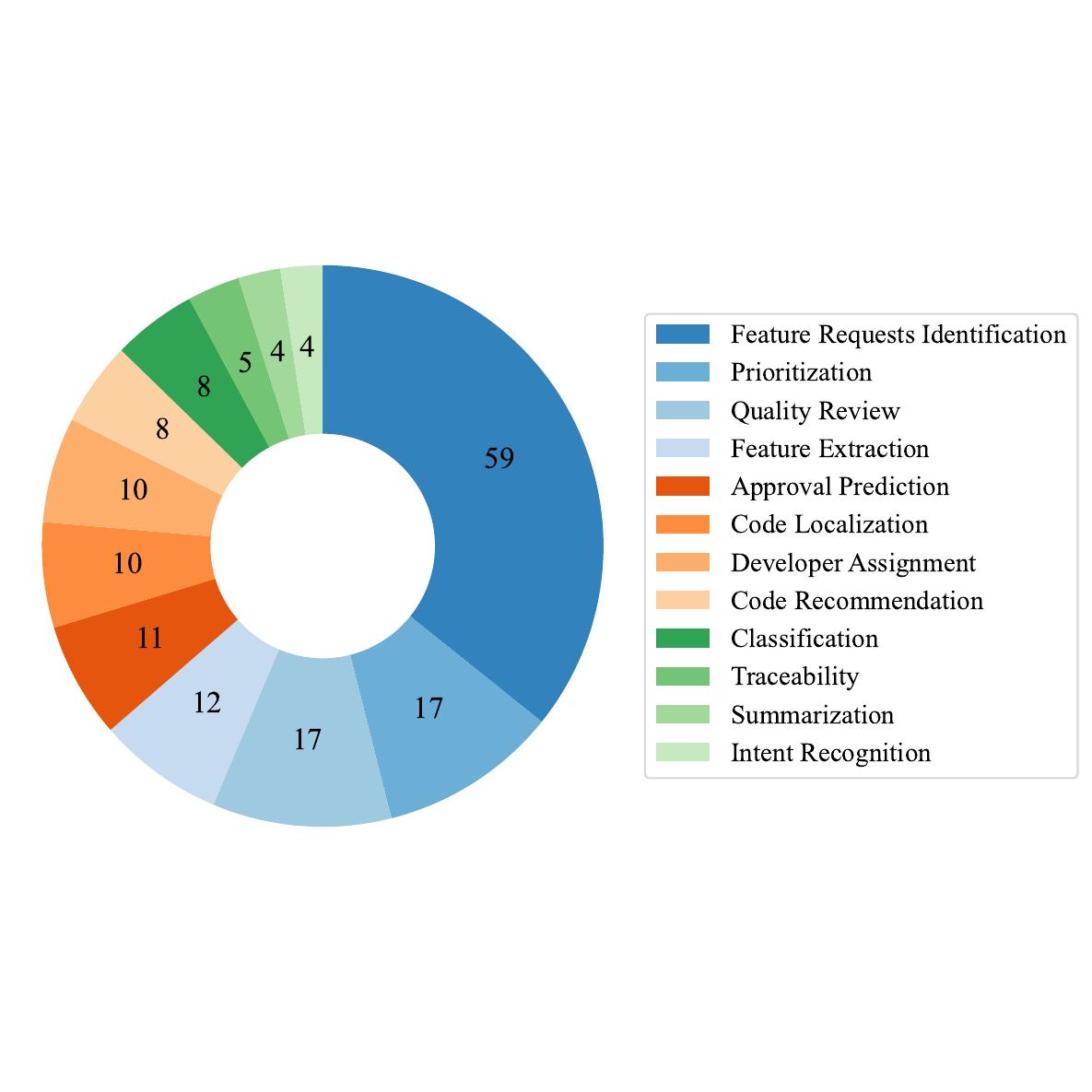}
    \caption{Research Topic Distribution.}
    \label{fig:topic}
  \end{minipage}%
  \hfill
  \begin{minipage}[t]{0.47\textwidth}
    \centering
    \includegraphics[width=\textwidth]{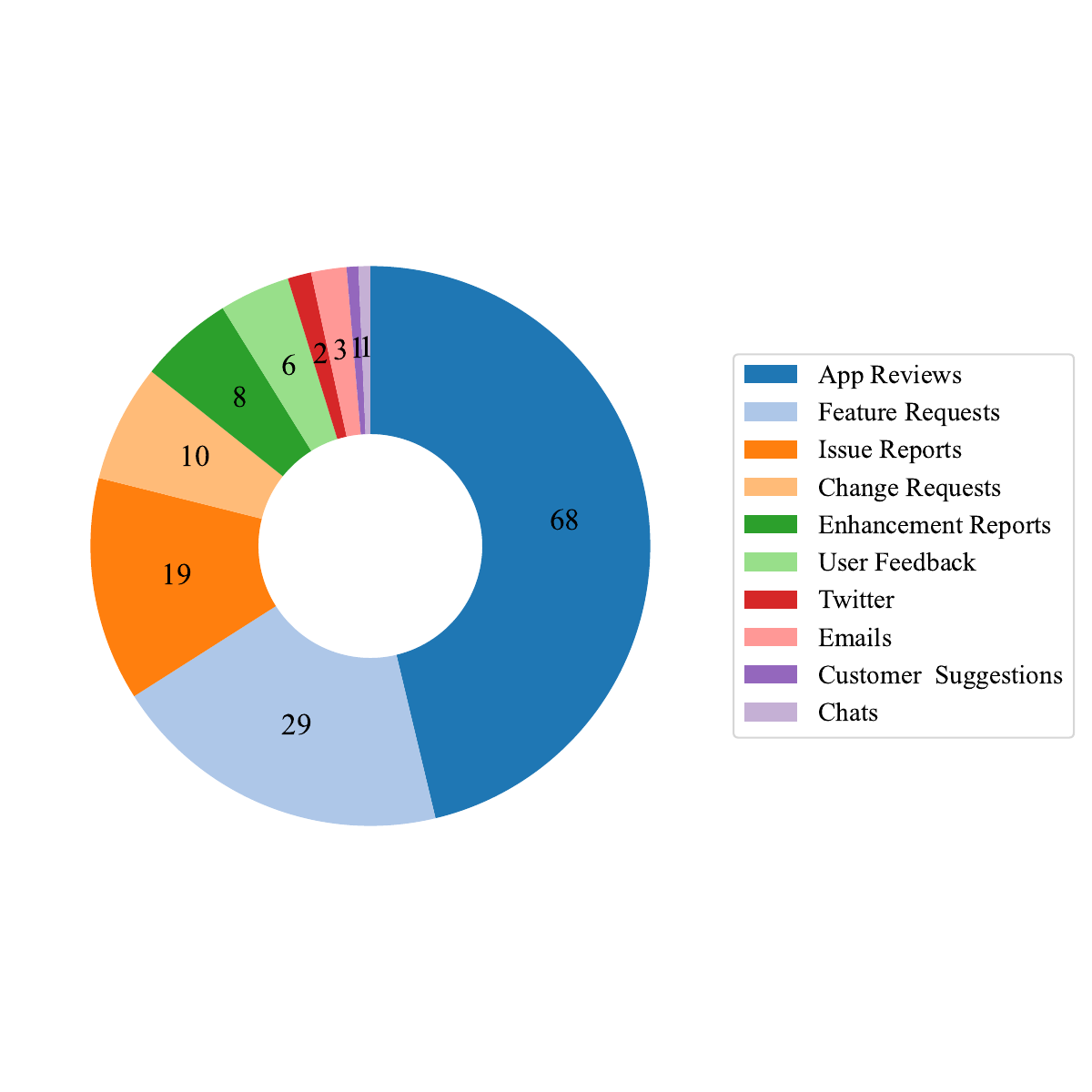}
    \caption{Source of Feature Requests.}
    \label{fig:source}
  \end{minipage}
\end{figure}

In terms of research focus, as illustrated in Figure~\ref{fig:topic}, twelve distinct research topics have been identified in the area of feature requests. The five most frequently studied topics are Feature Requests Identification (59), Prioritization (17), Quality Review (17), Feature Extraction (12), and Approval Prediction (11)\footnote{Some studies cover more than one topic.}. Together, these five topics account for over 70\% of the total studies, highlighting a strong emphasis on early-stage processing and decision-making. The remaining topics, while less frequently explored, suggest opportunities for future research to support downstream activities and enhance the overall lifecycle of feature request management.

\section{Analyzing and Processing Feature Requests} \label{sec:analysis}
Throughout the lifecycle of feature requests, twelve research topics have been identified in the literature on feature request analysis and processing: \textit{feature request identification}, \textit{quality review}, \textit{approval prediction}, \textit{feature extraction}, \textit{intent recognition}, \textit{summarization}, \textit{classification}, \textit{prioritization}, \textit{traceability}, \textit{developer assignment}, \textit{localization}, and \textit{code recommendation},
covering the entire process. 

In the following subsections, we present the state of the art across these twelve research topics, grouping and summarizing the relevant studies according to their respective focus areas.

\subsection{Feature Requests Identification} \label{sec:elicitation}

Feature requests are often embedded in diverse communication channels and platforms, including issue tracking systems, support forums, app store reviews, and social media comments. Figure~\ref{fig:source} illustrates the various forms of primary sources for feature requests as identified in existing research. As shown, 
app reviews (68) and issue reports (19) are the most frequently used sources for extracting feature requests.
Although user feedback varies in format across different platforms, it typically contains feature requests either explicitly or implicitly. For example, Iacob and Harrison~\cite{iacob2013retrieving} found that approximately 23.3\% of application reviews include feature requests---often in the form of user suggestions for new functionality or improvements to existing features. Similarly, Williams and Mahmoud~\cite{grant2017mining} reported that around 51\% of user comments on Twitter provide useful information, comprising 27\% bug reports and 24\% user requests. In addition to feature requests, user feedback may also include bug reports and insights into user experience. 

Feature request identification refers to the task of identifying and extracting descriptions of desired new software functionalities from such heterogeneous and often unstructured data. Given the sheer scale of user feedback—especially in commercial platforms such as the App Store, which hosted over 1.91 million apps, with an average of 1,342 new apps released every day and generated an immense volume of user reviews by early 2025~\footnote{\url{https://www.tekrevol.com/blogs/apple-app-store-statistics}}—manual extraction is infeasible for real-world applications, despite its potential accuracy.

To address this challenge, researchers have proposed a wide range of automated and semi-automated techniques, including unsupervised learning, supervised learning, semi-supervised learning, and hybrid methods. In addition, with the widespread adoption of large-scale pre-trained language models (PLMs), recent studies have introduced methods that leverage such models for feature request identification. These approaches range from fine-tuning task-specific models on labeled data to directly prompting general-purpose large language models (LLMs), such as GPT-3~\cite{brown2020language}, to extract or classify feature requests from user feedback. Table~\ref{tab:elicitation} summarizes representative methods used in the literature, highlighting the variety of strategies and learning paradigms designed to cope with the noise, diversity, and domain specificity of user feedback.

\begin{table*}[t!]
\caption{Classification of Approaches for Feature Requests Extraction.} \label{tab:elicitation}
\small
\begin{tabular}{c c p{7cm}}
\hline
\multicolumn{2}{ c }{Approaches} & Studies \\
\hline
\nff{Rule-based} & &
\cite{laura2013analysis}\cite{iacob2013retrieving}
\\ \hline
\multirow{6}{*}{Supervised}     & \begin{tabular}[t]{@{}c@{}}Text \\ Classification \end{tabular} &
\cite{abbas2024classification} \cite{al2021classification}
\cite{deeksha2019analysis} \cite{aslam2020convolutional} 
 \cite{dkabrowski2019finding} \cite{emitza2015ensemble} 
 \cite{izadi2022predicting} \cite{jha2017mining} \cite{kallis2019ticket} \cite{kallis2021predicting}  \cite{khan2019mining} \cite{javed2020conceptualising} \cite{khan2024mining} \cite{maalej2016automatic} \cite{walid2015bug} \cite{mcilroy2016analyzing} \cite{nafees2021machine} \cite{nayebi2018app} \cite{zhenlian2016an} \cite{lin2020detection} \cite{rui2020automatically} \cite{christoph2019classifying} \cite{tizard2019requirement} \cite{tong2018mining} \cite{lorenzo2016release} \cite{grant2017mining}  \cite{aman2020a} \cite{young2020comparison}
\\ \cline{2-3}
 &  \begin{tabular}[t]{@{}c@{}}Sentence \\ Classification \end{tabular} & \cite{di2016would} \cite{di2017surf} \cite{huang2018automating} \cite{li2020deep} \cite{thorsten2016software} \cite{panichella2016app} \cite{sebastiano2015how} \cite{faiz2019using} \cite{andrea2015development} \cite{james2019can}
 \\ \cline{2-3}
  &   Similarity  &\cite{tao2019labelling}
\\ \hline 
 Semi-supervised &  &
\cite{roger2017preliminary} \cite{venkatesh2018app} 
\\ \hline
 Pre-trained LM   &    & 
\cite{biswas2024interpretable} \cite{gambo2024enhancing} 
 \cite{hadi2023evaluating} \cite{motger2025leveraging} \cite{wang2023pros}
\\ \hline

 LLM  &   &
\cite{assi2024llm} \cite{ghosh2024exploring} \cite{ren2024combining} \cite{wei2024getting}
 \\ \hline
  Hybrid &     &  \cite{panichella2016app}\cite{lin2021automatically}      \\  
                         \hline
\end{tabular}
\end{table*}

\subsubsection{Unsupervised Learning-based Approaches}



Early approaches to identifying feature requests from user feedback mainly relied on unsupervised, rule-based extraction techniques. A typical method involved manually creating extraction rules, which enabled the identification and retrieval of feature request descriptions that matched predefined patterns. For instance, Iacob et al.~\cite{iacob2013retrieving} designed and implemented a prototype tool for mining feature requests from user reviews, defining 237 semantic extraction rules and employing automated matching algorithms to facilitate accurate extraction. Similarly, another study utilized the Aspect and Sentiment Unification Model (ASUM)—an information retrieval-based approach—to derive constructive user feedback from reviews, subsequently proposing these insights as new or improved software requirements for future versions~\cite{laura2013analysis}.

Rule-based matching methods offer high accuracy in extracting feature requests, but they also come with notable drawbacks. The reliance on manually curated rules results in substantial human effort, increasing cost and resource demands. Furthermore, manually formulated rules often fail to encompass the wide variety of scenarios present in real-world user feedback, thereby restricting the comprehensiveness and generalizability of these methods.

\subsubsection{Semi-supervised Learning-based Approaches}

Semi-supervised methods for feature request extraction typically encompass techniques such as self-training~\cite{david1995unsupervised}, Rasco~\cite{jiao2008a}, Rel-Rasco~\cite{yusuf2010cotraining}, and active learning~\cite{james2004active}. For example, Deocadez et al.~\cite{roger2017preliminary} employed three semi-supervised algorithms---self-training, Rasco, and Rel-Rasco—to categorize app reviews into bugs, requests, and other categories. They evaluated these methods using four base classifiers: \textit{K}-Nearest Neighbors (KNN), C4.5, SMO, and Naïve Bayes (NB). Their results demonstrated that semi-supervised approaches can substantially reduce the need for labeled data, achieving comparable classification accuracy using only 30\% of the labeled data required by fully supervised methods. Similarly, Dhinakaran et al.~\cite{venkatesh2018app} utilized active learning to extract feature requests from user reviews, applying three uncertainty sampling strategies---lowest confidence, margin sampling, and entropy---to identify the most informative samples for annotation. Their study found that active learning notably decreases labeling effort while maintaining performance on par with traditional supervised learning.

\subsubsection{Supervised Learning-based Approaches}

Supervised learning remains the most widely used approach for identifying feature requests from user feedback. Depending on granularity, methods operate at either the \textit{text} or \textit{sentence} level, labeling entries or individual sentences as feature requests, bug reports, or other categories~\cite{abbas2024classification, al2021classification, deeksha2019analysis}. 

\begin{figure}[t!]
  \centering
  \includegraphics[width=\textwidth]{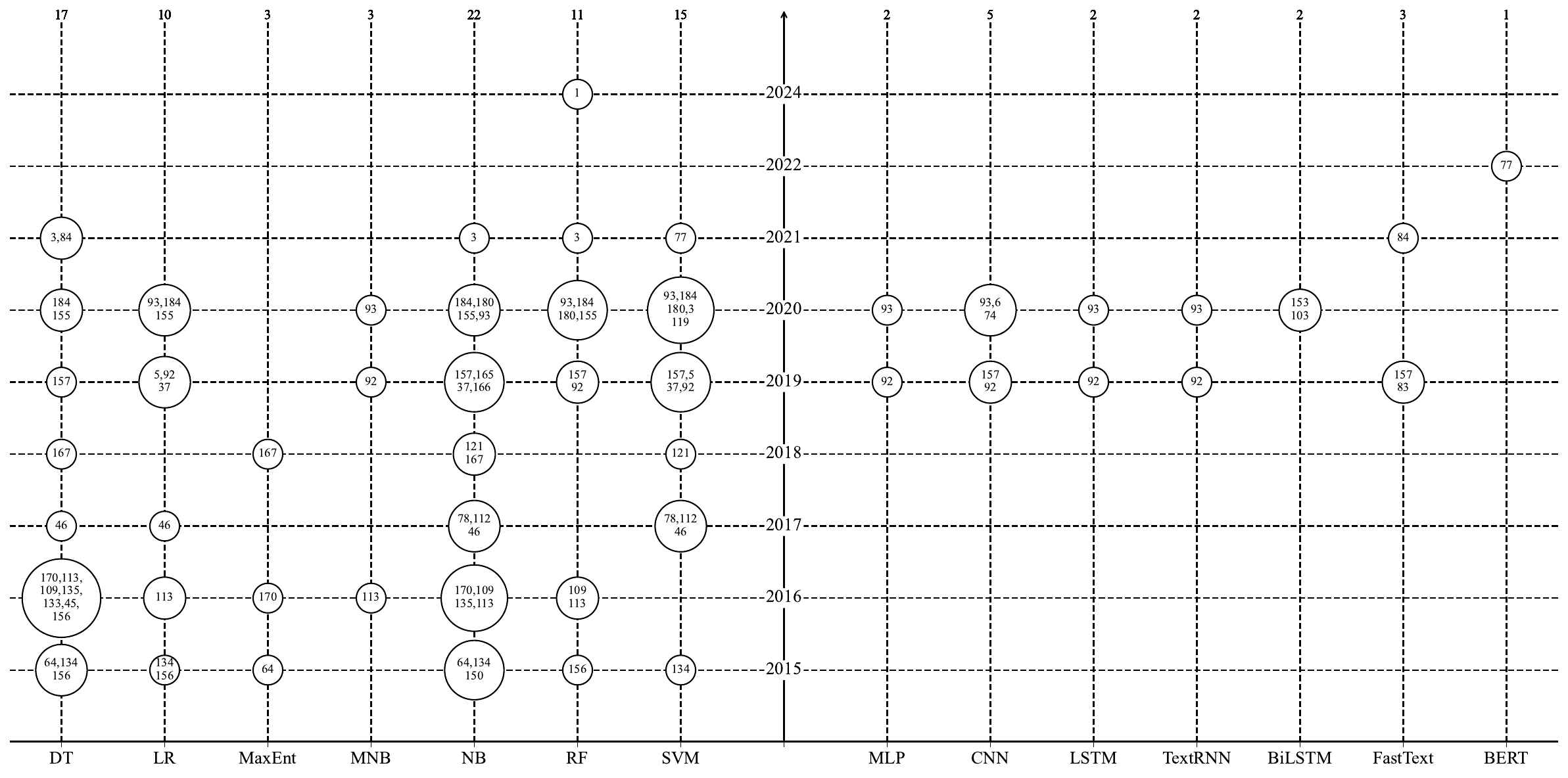}
  \caption{The Common Text Classification Algorithms.}
  \label{fig:common text classification algorithms}
\end{figure}

As shown in Figure~\ref{fig:common text classification algorithms}, traditional classifiers, such as NB and Decision Tree (DT), remain the most common, followed by Random Forest (RF) and Support Vector Machines (SVM). Deep learning methods began emerging around 2019, with Convolutional Neural Networks (CNN) becoming the most prevalent deep model for feature request extraction.
Performance varies depending on conditions and datasets. Guzman et al.~\cite{emitza2015ensemble} showed that ensemble classifiers outperform single classifiers. Stanik et al.~\cite{christoph2019classifying} observed that traditional classifiers excel on small datasets, whereas deep learning models perform better with large datasets. Khan et al.~\cite{khan2024mining} evaluated 14 algorithms, finding that MLP and RF achieve the highest average accuracy. In contrast, deep models like CNN, BiGRU, and GRU better capture complex textual patterns, excelling in precision, recall, and F-score for certain tasks. Ali Khan et al.~\cite{javed2020conceptualising} cautioned that deep learning may overfit on small samples, emphasizing appropriate classifier choice. To address imbalanced classes, Young et al.~\cite{young2020comparison} investigated resampling techniques (random undersampling, SMOTE oversampling, random oversampling, and SMOTE-Tomek) on five classifiers, finding that oversampling often yields better results. Nafees et al.~\cite{nafees2021machine} integrated semantic textual features and metadata into a gradient boosting machine (GBM) model with SMOTE-Tomek to predict approval probabilities for feature requests. Abbas et al.~\cite{abbas2024classification} combined RF with an accuracy sliding window (ASW) technique, improving classification accuracy while reducing tree counts, providing robust performance and new insights.

Beyond textual features, many studies incorporate auxiliary metadata such as review topics, rating levels, text length, sentence position, app size, installation counts, and reviewer numbers~\cite{emitza2015ensemble,walid2015bug,zhenlian2016an,deeksha2019analysis,aslam2020convolutional,thorsten2016software,james2019can}. Additional auxiliary features mined via NLP include keywords~\cite{walid2015bug,zhenlian2016an,rui2020automatically,thorsten2016software}, sentiment analysis~\cite{grant2017mining,emitza2015ensemble,walid2015bug,zhenlian2016an,aslam2020convolutional,aman2020a,sebastiano2015how,panichella2016app}, semantic rules~\cite{walid2015bug,rui2020automatically}, heuristic sentence identification~\cite{sebastiano2015how,panichella2016app,andrea2015development}, lexical features~\cite{panichella2016app}, and part-of-speech (POS) tagging~\cite{sebastiano2015how,panichella2016app,andrea2015development}.

\subsubsection{Pre-trained Language Model Approaches}

Mobile app reviews are often plagued by challenges such as low quality, grammatical errors, and subjective language. Traditional feature requests identification methods suffer from low recall rates, poor generalization, and difficulties handling noise, long text semantics, and cross-domain adaptation. Recently, PLMs have sparked significant advances in software requirements analysis by effectively integrating NLP techniques with requirements elicitation tasks.

Most PLM-based feature request identification methods build upon Transformer architectures and augment them with task-specific layers to perform classification or sequence labeling. For example, Hadi and Fard~\cite{hadi2023evaluating} evaluated BERT, RoBERTa, and ALBERT for classifying app reviews, demonstrating superior performance over traditional approaches like AR-Miner~\cite{chen2014ar-miner} and SUR-Miner~\cite{gu2015parts}across binary, multi-class, zero-shot, and multi-task scenarios. Their approach highlights the benefits of domain-adaptive pre-training in enhancing the semantic understanding of user reviews.

Despite their effectiveness in handling large-scale data, PLMs exhibit a black-box nature, making their decision processes opaque. This opacity hinders developers’ ability to fully trust and utilize model outputs, especially considering potential biases in the training data. To address these issues, recent research has emphasized the importance of interpretability and transparency. For instance, Gambo et al.~\cite{gambo2024enhancing} proposed a method that combines a BiLSTM with attention mechanisms and leverages local interpretable model-agnostic explanations (LIME) and SHapley Additive exPlanations (SHAP) to improve feature request identification. This approach outperforms SVM and CNN in accuracy and F1 score. Additionally, Biswas et al.~\cite{biswas2024interpretable} developed the IARCT method, which quantifies the contribution of each word via Shapley values and provides contextual explanations for Transformer-based classification decisions. 

In summary, PLM-based methods mark a promising direction for feature request identification, combining powerful semantic understanding with evolving interpretability techniques to bridge the gap between performance and explainability.

\subsubsection{General Large Language Model Approaches}
LLMs have significantly advanced the identification of feature requests from user reviews by leveraging their powerful natural language understanding capabilities to interpret unstructured feedback and identify critical requirements. Recent research has extensively explored the application of LLMs in analyzing app store reviews to enhance the accuracy of feature request identification, often complementing traditional machine learning methods through prompt-based task definition.

Ghosh et al.~\cite{ghosh2024exploring} evaluated the performance of \nff{two PLMs (BERT and DistilBERT) and one LLM (GEMMA)} on 3,200 Google Play Store user reviews. Their study confirmed the effectiveness of LLMs in parsing unstructured text and accurately identifying feature requests, showcasing the potential of LLMs to transform user feedback into actionable requirements.
Building on this, Assi et al.~\cite{assi2024llm} introduced the LLM-Cure framework, which derives feature improvement suggestions from user reviews via a two-phase approach. First, it extracts high-frequency functional features by semantically clustering batch reviews and normalizing cross-app features using word embeddings and cosine similarity. Then, for low-rated reviews of a target app, it retrieves corresponding high-rated competitor reviews to create contextual input for generating specific improvement suggestions through retrieval-augmented generation. 
Addressing challenges such as labor-intensive manual analysis and low clustering accuracy, Ren et al.~\cite{ren2024combining} proposed a hybrid method that combines Latent Dirichlet Allocation (LDA) topic modeling for user review classification with a GPT-4–based generation approach for categories with sparse reviews. 
Furthermore, Wei et al.~\cite{wei2024getting} conducted a comparative study of feature refinement approaches based on app store data and LLMs. Selecting 20 high-level root features (half existing and half novel) across diverse domains, they utilized App Store search and GPT-4 to generate two-level feature trees. 
Their integrated approach, implemented in the piStar~\cite{pimentel2018pistar} tool, reduces effort and inspires creativity, providing practical guidance for selecting appropriate feature request identification methods.

\subsubsection{Other Approaches}
Rahimi et al.~\cite{mona2014personas} proposed a semi-automated method for identifying feature requests from online forums by clustering user comments via incremental diffusion clustering and applying Apriori association rule mining to identify stakeholder-relevant topic clusters. 
Zhang et al.~\cite{tao2019labelling} developed a hybrid similarity-based approach to classify issue reports as bugs, feature requests, or others by combining normalized text similarities between unlabeled issue reports, user reviews, and labeled reports with weighted contributions.
Shi et al.~\cite{lin2020detection} introduced FRMiner, which uses a deep Siamese network to determine whether chat messages belong to the same class by learning their similarity, enabling feature requests identification.
Panichella et al.~\cite{sebastiano2015how} combined manual rule matching of 246 linguistic patterns with sentiment analysis and J48 classification to categorize user reviews into information provision, search, feature requests, problem detection, and others.
Gribkov et al.~\cite{gribkov2020neural} presented the Trans-LSTM model to extract (topic, description) expressions from user review sentences, classifying them into bug reports, feature requests, and positive or negative features. Their use of Trans-CNN and Trans-LSTM outperformed prior models.
Shi et al.~\cite{lin2021automatically} proposed an automated method for identifying feature requests from development emails using semantic sequence patterns generated from fuzzy rules and email context represented via a 2-gram model, validated on Ubuntu community emails.
Izadi et al.~\cite{izadi2022predicting} introduced a multi-task approach that predicts both issue type (bug, feature, question) using fine-tuned BERT, compared with RF and SVM, and issue priority (high, medium, low), incorporating the predicted issue type to enhance cross-task learning.

\textcolor{black}{Feature identification focuses on transforming unstructured user feedback into explicit and actionable feature requests that guide subsequent localization and implementation. 
Nadeem et al.~\cite{nadeem2021extracting} introduced a labeled dataset and evaluated supervised learning models, including LR, DT, and SVM, to identify software change request sentences in app reviews, demonstrating high effectiveness across models. 
Wu et al.~\cite{wu2021identifying} proposed KEFE, which extracts feature-describing phrases from app descriptions, aligns them with semantically related user reviews using a deep classifier, and applies regression analysis to measure their correlation with app ratings. 
AI-Safoury et al.~\cite{al2022integrating} integrated app-store reviews with Github issue reports into unified change request lists through syntactic and sentiment-based classification, similarity-driven deduplication, and prioritization. 
Finally, Cai et al.~\cite{cai2025automatic} provided a systematic review of user-generated-content-based requirements elicitation, outlining how rule-based, topic modeling, and machine learning techniques have converged into a data-driven framework for automated feature identification.}

\subsection{Quality Review}
Quality review is a critical activity in RE, aimed at assessing the quality, correctness, and feasibility of incoming requirements. As noted by Wiegers~\cite{wiegers2003software}, a comprehensive review should examine aspects such as the organization, completeness, correctness, quality attributes, and traceability of the requirements specification. Functioning as a ``filter'' within the requirements analysis process, feature request review helps eliminate low-quality requests—such as those that are ambiguous, redundant, or irrelevant—before they reach the development phase, thereby reducing errors and defects downstream~\cite{wiegers2003software}.

Existing research on feature request review primarily centers on defining quality standards~\cite{heck2015quality,shi2016new} and employing these criteria to assess the quality of requests~\cite{heck2017framework,heck2013analysis}. In addition, a growing body of work has proposed automated techniques to detect duplicate~\cite{aversano2019issue,gu2011analysis, li2017detecting,heck2014horizontal}, redundant~\cite{shi2016new}, and ambiguous~\cite{aslam2020convolutional} feature requests, aiming to streamline the review process and improve the overall quality of requirements.

Heck and Zaidman~\cite{heck2015quality, heck2017framework} developed a framework for assessing just-in-time (JIT) requirements, emphasizing completeness, uniformity, and correctness as core criteria. Aversano~\cite{aversano2019issue} conducted empirical studies on enterprise open-source systems to examine whether issue reports include detailed entries, attachments, comments, and maintain readability—all indicative of quality. In an investigation of rejection reasons in open source projects, Shi et al.~\cite{shi2016new} identified redundancy (23.37\%), duplication (18.90\%), conflict (4.67\%), and incompleteness (3.86\%) as primary causes. Furthermore, duplication has been found to be particularly prevalent, with studies reporting rates as high as 36\% in Mozilla Firefox’s feature requests~\cite{heck2013analysis}. This has motivated a body of research on automatically identifying duplicates and redundancies. Shi et al.~\cite{shi2016new} proposed an approach that combines cosine similarity between textual vectors with structural similarity analysis based on software feature trees constructed from user manuals and release notes to identify redundant software features. Gu et al.~\cite{gu2011analysis} have applied unsupervised clustering based on TF-IDF to recommend potentially duplicated reports at submission time. Similarly, Li et al.~\cite{li2017detecting} used cosine similarity between the summaries and descriptions of new and existing requests to generate candidate duplicates, using a predefined threshold for filtering. Heck and Zaidman~\cite{heck2014horizontal} summarized causes of duplication—including repeated solutions, partial overlaps, patches submitted for existing requests, and varied terminology. While existing approaches have achieved promising recall rates, challenges remain in detecting more nuanced cases, such as partial overlaps or requests phrased differently. Ambiguity also poses a significant obstacle, as identified by Gill et al.~\cite{kanwal2014semi}, who categorized natural language ambiguity into lexical, syntactic, and semantic and recommended combining domain-specific knowledge with automation to mitigate misclassification. More recently, Ghandiparsi et al.~\cite{ghandiparsi2025towards, ghandiparsi2025demystifying} investigated GPT‑4o’s ability to detect ambiguity and incompleteness in feature requests, and to generate clarifying questions that help refine these requests in open‑source software development contexts.  

Collectively, these studies underscore the importance of quality evaluation in feature request reviews and suggest both manual and automated approaches for enhancing review effectiveness and minimizing redundancy in requirements management.

\subsection{Approval Prediction}
Feature request approval prediction (FRAP) involves evaluating the necessity and feasibility of feature requests to determine if they should be approved and developed. Timely identification and prioritization of valuable user feedback significantly influence software innovation~\cite{erne2020investigating}. 

Early research employs machine learning classification models for automatic FRAP: Nizamani et al.~\cite{Nizamani2018automatic} developed a MNB classifier, achieving an accuracy of 89.25\%, outperforming models like NB, SVM, RF, and LR. Umer et al.~\cite{umer2019sentiment} enhanced this model by integrating sentiment analysis results from Senti4SD, which significantly improved predictive accuracy.
Arshad et al.~\cite{arshad2021deep} introduced a deep learning approach combining word2vec embeddings and sentiment analysis to capture syntactic and semantic information in feature requests, achieving superior performance compared to earlier methods by Nizamani et al.~\cite{Nizamani2018automatic} and Umer et al.~\cite{umer2019sentiment}. Nafees et al.~\cite{nafees2021machine} compared SVM, MNB, and LR classifiers using a dataset of 40,000 enhancement reports from Bugzilla. Their study demonstrated that the SVM classifier outperformed MNB and LR, achieving an accuracy of 90.38\% and an F-measure of 74.68\%. This highlighted SVM’s effectiveness in handling high-dimensional feature spaces and identifying critical textual features.

Fitzgerald et al.~\cite{fitzgerald2011early, fitzgerald2011early_extended} proposed predicting software faults by analyzing 13 distinct feature request attributes using DT, NB, LR, and M5P-tree algorithms. Their findings highlighted that BOW and TF-IDF vectorization effectively predict fault-prone requests, emphasizing the value of textual content.

Studies based on Bugzilla data demonstrated that sentiment analysis notably boosts prediction accuracy~\cite{Nizamani2018automatic, umer2019sentiment, arshad2021deep, nafees2021machine}. However, research using Google Play comments indicated limited effectiveness in predicting request implementation based on negative sentiments.

Moreover, Niu et al.~\cite{niu2022towards} observed temporal dependencies among feature requests, recommending chronological data splitting over traditional cross-validation to avoid biased evaluation.

Heppler et al.~\cite{heppler2016who} revealed that enhancement requests submitted by developers were twice as likely to succeed compared to external submissions. Building on this insight, Niu et al.~\cite{niu2023utilizing} introduced a FRAP method that incorporates creator profiles (including ID, role as developer or not, and submission frequency), combined with TextCNN, which significantly enhances predictive performance. More recently, Zuo et al.~\cite{zuo2025enhancement} improved FRAP accuracy by integrating generative LLM (e.g., GPT-4o, DeepSeek-V3) with creator profiles.

\subsection{Feature Extraction}

A software feature is defined by Kang et al.~\cite{kang1990feature} as ``a prominent or distinctive user-visible aspect, quality, or characteristic of a software system or systems.'' Software feature extraction refers to the process of identifying and retrieving these aspects, qualities, or characteristics from software requirements using text mining techniques. According to Bakar et al.~\cite{bakar2016extracting}, extracted features can be provided to domain analysts as early indicators for related product development or used to construct software feature models. Additionally, extracting features from feature requests plays a crucial role in understanding user needs and uncovering dependencies among different features.

Bakar et al.~\cite{bakar2015feature} conducted a comprehensive survey of research efforts focused on extracting software features from natural language requirements prior to 2015. Their review summarized methods for extracting features from various natural language sources, including requirement specifications and product descriptions.

This section primarily investigates methods for extracting software features that users expect from user feedback and feature requests. Existing approaches for software feature extraction can be broadly categorized into three technical paradigms: manual annotation, lexical analysis-based methods, and language model-driven techniques.

\subsubsection{Lexical Analysis-Based Approaches}

Lexical analysis-based methods extract software features by identifying and filtering words or phrases that match predefined POS patterns, such as nouns, verbs, adjectives, or their combinations, to capture meaningful feature expressions from textual data. Based on these filtering rules, candidate software features are selected accordingly. 


Bakar et al.~\cite{bakar2016extracting} introduced an approach based on expert review texts, which preprocesses the data by selecting nouns, verbs, and adjectives, and then removes marginal documents through clustering. 
Similarly, Bakiu et al.~\cite{elsa2017which} employed a collocation algorithm on user reviews, filtering for core POS tags and selecting features based on frequency and proximity, combined with lexical sentiment analysis to assess user satisfaction with the extracted features. Johann et al.~\cite{timo2017safe} proposed the SAFE method, which relies on 18 manually defined POS patterns and five sentence structures to extract software features from both application descriptions and user reviews. Notably, SAFE requires minimal training data and has been widely adopted~\cite{faiz2019using, shah2019safe}.

Regarding priority ranking and competitive analysis, Licorish et al.~\cite{licorish2017attributes} focused on noun extraction from reviews to predict the processing priority of feature requests, while Dalpiaz et al.~\cite{dalpiaz2019re} utilized POS tagging and various POS combinations, such as $<noun+adjective>$ and $<adjective+verb>$, to mine features from competitor app reviews. 
Shah et al.~\cite{faiz2018the} proposed annotation guidelines emphasizing the labeling of only consecutive word phrases containing nouns, limited to a maximum of three words, while excluding references to the application itself.

\subsubsection{Language Model-Driven Techniques}
Language model-driven techniques exploit the powerful contextual understanding of PLMs, particularly Transformer-based LLMs, to improve software feature extraction from unstructured user feedback. These approaches typically formulate feature extraction as token classification or generation tasks, leveraging semantic representation to overcome the limitations of traditional lexical analysis-based approaches.
Cong et al.~\cite{cong2023small} developed ISIFRank, an enhanced version of SIFRank that integrates the ERNIE pre-trained model with innovative POS combination rules. This approach effectively extracts key product information phrases reflecting core user needs, outperforming traditional keyword extraction methods in both precision and F1-score. 
Gambo et al.~\cite{gambo2024enhancing} combined Bidirectional Long Short-Term Memory (BiLSTM) networks with attention mechanisms, leveraging n-gram features, TF-IDF encoding, and pre-trained word embeddings to capture semantic and contextual patterns from unstructured user feedback. Their method incorporates named entity recognition and POS tagging to accurately identify software features, demonstrating the synergy of deep learning and traditional NLP techniques.

Motger et al.~\cite{motger2025leveraging} pioneered the use of encoder-based LLMs (e.g., BERT, RoBERTa, XLNet) for feature extraction from app reviews by framing it as a supervised token classification task enhanced with domain-adaptive pre-training and instance selection to improve accuracy and efficiency. This approach significantly outperforms earlier syntactic pattern methods by effectively handling noisy and low-quality reviews.
Building on this, Wei et al.~\cite{wei2024getting} demonstrated two complementary strategies: one that leverages app store metadata with embedding-based search to extract sub-features, and another that employs GPT-4 to generate hierarchical feature trees directly from textual descriptions. 
Zhang et al.~\cite{zhang2024user} tackled the challenge of terminology mismatch by integrating semantic similarity computations with BERT embeddings, aligning informal user language with formal product attribute vocabularies. This improves the precision of feature attribute extraction within product design contexts.

Together, these studies illustrate a trend toward increasingly sophisticated use of LLMs to capture semantic nuances, handle domain-specific language, and produce structured, reliable feature representations from noisy natural language input.

In summary, lexical analysis approaches rely on predefined POS patterns and syntactic rules to identify candidate feature phrases, offering a straightforward and interpretable solution that works well with limited data and requires minimal training. However, they often struggle with semantic ambiguity, noise, and linguistic variability inherent in informal user reviews.
In contrast, language model-driven approaches leverage the powerful contextual and semantic understanding capabilities of PLMs, such as BERT and the GPT series. These methods can effectively handle noisy, unstructured, and domain-specific language, enabling more accurate and flexible extraction of complex and hierarchical feature representations. While more computationally intensive, language model-driven techniques have demonstrated superior performance, especially when combined with domain adaptation, semantic similarity measures, and generation-based frameworks.

\subsection{Intent Recognition}
User intent recognition plays a crucial role in understanding the underlying goals and motivations behind feature requests submitted by end-users. When users communicate their needs to developers, they often provide contextual explanations, reasons, and even suggested solutions to make their requests clearer and more actionable. Accurately mining these intents from unstructured user feedback enables requirements analysts and developers to understand user priorities better, leading to more effective requirement specification and prioritization.

Early approaches combined NLP with fuzzy rule-based systems that analyze linguistic features at vocabulary, syntax, and semantic levels to identify sentence patterns corresponding to different intents such as explanation, advantage, disadvantage, and example. Integrating these fuzzy rules with classifiers such as RF, NB, and SVM significantly improves classification performance~\cite{lin2017understanding}. Building on this, tools like NERO~\cite{mu2020nero} automate the classification of sentences within feature requests using such rule-based techniques.
Beyond rule-based systems, methods grounded in speech-act theory classify user feedback at the sentence level into fine-grained intents, including requestive, suggestive, and assertive types. By aggregating these intent signals, developers gain clearer insights into user demands, aiding prioritization efforts~\cite{kifetew2021automating}.
DT–based classifiers have also been applied to detect specific intents, such as update requests in developer forums like Stack Overflow. Such studies reveal that a significant portion of user comments contain update requests, many of which receive timely responses, underscoring the practical importance of intent recognition in community feedback~\cite{sheikhaei2023study}.


In summary, user intent recognition has evolved from rule-based and traditional machine learning methods to sophisticated, theory-informed approaches, and most recently, interactive LLM-powered systems. Together, these methods enhance the understanding of user motivations in feature requests, facilitating more accurate requirement elicitation, prioritization, and ultimately improving software responsiveness to user needs.

\subsection{Summarization}

Summarization is another task designed to enhance the understanding of users' intent. Unlike the structured narrative format of user stories, feature requests are often expressed informally and colloquially, making them challenging to interpret. To overcome this, various summarization techniques have been developed to extract core topics and standardize the language of feature requests, thereby enhancing their clarity and usability.

Early work by Jha et al.~\cite{nishant2018using} introduced MARC 2.0, which employs frame semantics via FrameNet for semantic role labeling to annotate app store user reviews. Following this, summarization algorithms---including Hybrid TF, Hybrid TF-IDF, SumBasic, and LexRank---were applied to generate summaries for defect reports and feature requests. Their evaluation demonstrated that the SumBasic algorithm produced summaries closely aligned with manual summaries, effectively capturing essential information.
Similarly, Williams et al.~\cite{grant2017mining} focused on classifying software-related tweets into bug reports and user requirements, using the same summarization techniques. The redundancy control inherent in SumBasic enabled occasional inclusion of low-frequency words, making its summaries particularly faithful to human-generated ones.
Gao et al.~\cite{gao2022listening} introduced the SOLAR framework, which automatically summarizes user reviews by combining review helpfulness prediction, topic-sentiment modeling (using Biterm Topic Model and Binary Sentiment Topic models), and a multi-factor ranking module. This approach clusters topics, analyzes sentiment, and ranks reviews based on semantic representativeness and sentiment priority, allowing developers to identify core user feedback quickly.
More recent advances incorporate deeper semantic and argumentative structures. Wang et al.~\cite{wang2023pros} proposed a graph-ranking stance summarization method that integrates semantic relevance between comment sentences and feature descriptions with argumentative roles (e.g., major claims, premises). This approach constructs a graph where sentence centrality is influenced by both semantic similarity and argumentative function, enabling the extraction of balanced summaries that support or oppose particular stances. Their VOTEBOT system achieved high ROUGE-1 scores, demonstrating superior performance over traditional summarization methods and offering the first structured representation of bipolar opinion in feature request discussions.

This evolution from simple frequency-based summarization to sophisticated, semantically rich, and argument-aware methods highlights the ongoing effort to improve the interpretability and usability of feature request summarization in software engineering.

\subsection{Classification}

Requirements classification is a fundamental activity within requirements analysis that involves categorizing software requirements according to predefined classification dimensions and attributes. Unlike the classification algorithms used during feature request identification, classification here focuses on organizing a large volume of feature requests that span diverse aspects of software functionality and quality attributes such as security, reliability, and usability. Proper categorization not only facilitates more effective analysis by requirements engineers but also supports the structured storage and management of user feature requests.

Existing research on feature request classification can be broadly examined from three perspectives: classification methods, features used for classification, and classification categories.

Table~\ref{ClassificationCategories} summarizes the various classification categories and methods adopted across current studies. While classification schemes vary, most differentiate between functional and non-functional requirements, further detailing non-functional characteristics. From a methodological standpoint, classification approaches fall into four main groups: supervised learning-based approaches~\cite{simone2019listening, li2018automatically}, unsupervised learning-based approaches~\cite{lorenzo2016release, cleland2009automated}, keyword co-occurrence matrix-based approaches~\cite{tao2020identifying}, deep learning-based approaches~\cite{niu2021deep}, and generative LLM-based approaches~\cite{bai2024prioritizing}.

\noindent \textbf{\ding{46} Classification Methods.} Supervised approaches predominantly utilize traditional machine learning classifiers such as KNN~\cite{li2018automatically, wang2018can}, NB~\cite{li2018automatically, wang2018can}, SVM~\cite{li2018automatically}, RF~\cite{simone2019listening}, J48~\cite{wang2018can}, and Bagging ensembles~\cite{wang2018can}. Conversely, unsupervised techniques primarily employ clustering algorithms, such as DBSCAN~\cite{lorenzo2016release, simone2019listening} and Spherical K-means~\cite{cleland2009automated}. Initial methodologies for requirement categorization favored unsupervised clustering techniques, predominantly before 2016. However, more recent work has increasingly gravitated toward deep learning-based and LLM-based approaches. Among these, deep learning methods typically employ several key architectures, including CNNs, LSTMs, BiLSTMs, GRUs, and BiGRUs~\cite{niu2021deep}. Concurrently, LLM-based approaches demonstrate significant effectiveness in this domain. Notably, these LLM approaches distinguish themselves by highlighting their specialized capabilities in natural language understanding for the specific task of requirement categorization~\cite{bai2024prioritizing}.


\noindent \textbf{\ding{46} Classification Features.} In addition to textual vectorization, many studies incorporate auxiliary features to enhance classification performance. Common supplementary features include project-specific keywords and heuristic attributes~\cite{li2018automatically, niu2021deep}, meta-data from user reviews such as ratings, review length, and application categories~\cite{simone2019listening}, and change logs~\cite{wang2018can}. Notably, project keywords have proven particularly effective, as feature requests describing similar performance levels tend to share common keyword sets~\cite{li2018automatically}.

\noindent \textbf{\ding{46} Classification Categories.} Regarding classification categories, feature requests are typically organized according to requirement levels described in the requests, such as security, usability, ease of use, among others. RE commonly distinguishes between functional requirements---detailing software functionalities---and non-functional requirements, which specify quality attributes essential for fulfilling user and business needs. Standards like ISO 25010 categorize software quality into functional and non-functional characteristics, with the latter encompassing functional suitability, performance efficiency, compatibility, usability, reliability, security, maintainability, and portability~\cite{ISO9126-1:2001}. Various studies have classified feature requests accordingly. Li et al.~\cite{li2018automatically} categorized requests into security, reliability, performance, lifecycle, usability, ease of use, and system interface. Niu et al.~\cite{niu2021deep} adopted the same dataset as Li et al. Scalabrino et al.~\cite{simone2019listening} categorized user comments into defect reports, functional requests, performance requests, security requests, energy consumption requests, availability requests, and other categories. Tao et al.~\cite{tao2020identifying} classified security-related comments into system, finance, spam, privacy, and others. Wang et al.~\cite{wang2018can} grouped requirements from user reviews into usability, reliability, portability, and performance. More recently, Bai et al.~\cite{bai2024prioritizing} used structural topic modeling to identify eight key user concerns from video conferencing app reviews, including data privacy, audio/video quality, customer service, security issues, meeting management, feature requests, and platform compatibility.

In conclusion, feature request classification is a crucial step in requirements analysis that facilitates effective organization, understanding, and management of diverse user requests. The field has evolved from early unsupervised clustering techniques to more advanced supervised learning and explainable AI approaches, leveraging both textual and auxiliary features for improved accuracy. Categorization schemes generally align with established RE frameworks, distinguishing functional and non-functional aspects and capturing detailed quality attributes. Continued advancements in classification methodologies and feature representation promise to further enhance automated analysis and prioritization of feature requests in software development.

{\small
\begin{table*}[h]
    \caption{Classification Categories in Existing Research.}
    \label{ClassificationCategories}
    \centering
    \begin{tabular}{m{0.07\textwidth}<{\raggedright}m{0.5\textwidth}<{\raggedright}m{0.25\textwidth}<{\raggedright}m{0.07\textwidth}<{\raggedright}}
        \toprule
        Paper & Category & Method & Year \\
        \midrule
        {\cite{cleland2009automated}} & Clustering based on software features & Spherical K - means clustering & 2009 \\
        \rowcolor[HTML]{EFEFEF} 
        {\cite{lorenzo2016release}} & Unsupervised clustering & DBSCAN clustering & 2016 \\
        {\cite{simone2019listening}} & Defect reports, function requests, performance requests, security requests, energy consumption requests, usability requests, others & RF & 2017 \\
        \rowcolor[HTML]{EFEFEF} 
        {\cite{li2018automatically}} & Security, reliability, performance, life cycle, usability, ease of use, system interface & KNN, NB, and SVM & 2018 \\
        {\cite{wang2018can}} & Usability, reliability, portability, performance & NB, Bagging, DT, and KNN & 2018 \\
        \rowcolor[HTML]{EFEFEF} 
        {\cite{tao2020identifying}} & System, funds, spam, privacy, others & Word co-occurrence matrix based on keywords & 2020 \\
        {\cite{niu2021deep}} & Security, reliability, performance, life cycle, usability, ease of use, system interface &  Deep learning models, Word2vec, TF-IDF, Keyword Frequency, Heuristic Properties  & 2021 \\
        \rowcolor[HTML]{EFEFEF} 
        {\cite{bai2024prioritizing}} & Steal data, audio and video quality, customer service, hacker issues, meeting and account passwords, mute and unmute, features, office platform & structural topic modeling with explainable AI & 2023 \\
        \bottomrule
    \end{tabular}
\end{table*}}

\subsection{Prioritization}
Prioritization of feature requests is a critical activity in software engineering, as it directly influences the planning, resource allocation, and timely delivery of software products. Given the typically large volume of user-submitted feature requests, effective prioritization ensures that development efforts focus on features that maximize customer value, enhance system quality, and align with business goals~\cite{lehtola2004requirements, achimugu2014systematic}. Poor prioritization can lead to wasted resources, delayed releases, and reduced user satisfaction. Consequently, feature request prioritization serves as a bridge between user needs and engineering constraints, facilitating informed decision-making throughout the software development lifecycle~\cite{chung2012non}.

Prioritization typically involves ranking requirements or features based on specific criteria---such as importance, cost, or risk---to determine which items should be included in the next software release. Common traditional methods for requirements prioritization include the Analytic Hierarchy Process (AHP)~\cite{karayalcin1982analytic}, Kano model analysis~\cite{1984KJ00002952366}, binary search tree (BST)~\cite{karlsson1998evaluation}, cumulative voting (CV)~\cite{leffingwell2000managing}, planning game (PG)~\cite{beck2000extreme}, MoSCoW~\cite{tudor2006using}, case-based ranking (CBRank)~\cite{avesani2005facing}, and priority group methods~\cite{sillitti2005requirements}. 
Notably, methods such as Kano and CBRank depend heavily on gathering user input to assess requirements. Conversely, approaches like AHP and CV involve computational complexities that grow rapidly with the size of the dataset, making them less practical for prioritizing extensive sets of feature requests.

Unlike traditional requirements, feature requests typically occur in much larger volumes and greater diversity, requiring prioritization strategies specifically designed to manage this scale and complexity. As a result, prioritization methods have evolved from straightforward user-driven approaches to advanced automated frameworks that integrate NLP, machine learning, and stakeholder input.

\noindent \textbf{\ding{46} User Voting and Contribution-Based Approaches.}
In vendor-driven open-source projects, prioritization often involves direct user participation. Users vote on feature requests, with higher votes indicating higher priority. In some virtual world games, stakeholders’ priority weight correlates with their contribution level. Meanwhile, in forums lacking explicit priority mechanisms, user comments implicitly determine priorities~\cite{laurent2009lessons}. These approaches leverage community input to guide prioritization but may lack scalability and fine-grained analysis.

\noindent  \textbf{\ding{46} Attribute-Based and Textual Analysis Methods.}
Several studies have focused on extracting meaningful attributes from user feedback to inform prioritization. Keertipati et al.~\cite{keertipati2016approaches} identified key features including request frequency, user ratings, negative sentiment, and modal verb usage, and developed ranking methods that balance developer expertise and data-driven insights via independent attribute evaluation, weighted scoring, and regression models. Licorish et al.~\cite{licorish2017attributes} similarly employed linguistic and sentiment features combined with regression models to classify user reviews into priority groups, enabling structured triage of issues.

\noindent  \textbf{\ding{46} Clustering and Machine Learning-Based Hybrid Approaches.}
To manage large-scale data, clustering combined with machine learning has become prevalent. AR-Miner~\cite{chen2014ar-miner} groups user comments using LDA and ASUM, ranking clusters based on comment counts, period, and scores. CLAP~\cite{lorenzo2016release} extends this by classifying comments via RF and clustering with DBSCAN, extracting cluster-level features for priority labeling. An improved CLAP~\cite{simone2019listening} further specializes in non-functional issues, showing better performance than AR-Miner.
Izadi et al.~\cite{izadi2022predicting} developed a priority prediction model that leverages TF-IDF, semantic tags, and normalized features (e.g., discussion length, developer reputation), achieving robust accuracy across projects. Kifetew et al.~\cite{kifetew2021automating} proposed ReFeed, which integrates semantic similarity with sentiment and intent weighting to automate prioritization, supported by domain ontologies and WordNet.
Etaiwi et al.~\cite{etaiwi2020order} combined CLAP-based clustering with consensus algorithms to derive optimized priority rankings, integrating custom review metrics to select features for development iterations.

\noindent  \textbf{\ding{46} Integration of Strategic Decision-Makers and Multi-Criteria Models.}
Morales et al.~\cite{morales2017exploiting} combined user feedback attributes with strategic decision-maker preferences to calculate requirement priorities, leveraging NLP techniques including POS tagging, sentiment analysis, and topic modeling to extract features such as sentiment polarity and severity. Malgaonkar et al.~\cite{svensson2024not} applied Bayesian regression to large industrial datasets to quantify the dynamic influence of criteria such as team priority, criticality, and business value across development stages, also proposing heuristic and regression-optimized prioritization models with high accuracy and efficiency.
Zhang et al.~\cite{zhang2024user} addressed the evolution of requirements by analyzing demand category transitions using the FSP-Kano model to guide product iteration through hierarchical priority principles. Salleh et al.~\cite{salleh2024review} developed a four-element prioritization framework integrating elicitation, classification, auxiliary factors, and priority scoring to support iterative planning under constraints.

\noindent  \textbf{\ding{46} Advanced Priority Marking and Risk-Based Approaches.}
Kegel et al.~\cite{kegel2023automating} introduced explicit priority keywords, utilizing modal verbs to distinguish between mandatory and optional requests, thereby enforcing a strict priority order that is executable at runtime. De et al.~\cite{de2024opinion} designed MApp-IDEA, which dynamically prioritizes issues by constructing a risk matrix from sentiment, clustering, and graph-based similarity metrics powered by LLMs, detecting risk spikes well ahead of software updates.

Feature request prioritization methods can be grouped into grouping methods (e.g., Kano, MoSCoW, priority groups), which assign requests into discrete priority levels~\cite{lorenzo2016release, li2018automatically, chen2014ar-miner}, and non-grouping methods that produce ranked lists through pairwise comparisons or weighted scoring schemes (e.g., AHP, BST, CBRank, planning game, cumulative voting)~\cite{keertipati2016approaches, morales2017exploiting, etaiwi2020order}. Grouping methods provide intuitive categorizations but often require extensive human input and may not scale well, while non-grouping approaches offer fine-grained rankings but can be computationally costly.

Our key observations include:
\ding{192} Automated prioritization is essential due to the large, unpredictable volume of feature requests, significantly reducing manual effort.
\ding{193} Extracting rich textual and meta-attributes (frequency, sentiment, ratings, clustering features) improves prioritization accuracy.
\ding{194} There is a strong correlation between user request volume and developer response frequency, validating user demand as a priority indicator~\cite{licorish2015satisfying}.
While user-centric indicators aid in identifying popular demands, they risk neglecting critical but less visible features and often overlook development constraints, such as cost and time. An effective prioritization framework should strike a balance among user importance, stakeholder constraints, and expected benefits.

Future research should advance automated, multi-factor prioritization models integrating user feedback, project feasibility, and cost-benefit trade-offs to optimize feature implementation in evolving software systems.

\subsection{Traceability}
Traceability refers to the ability to establish and examine relationships between data stored in different artifacts throughout the software development process. Gotel defines traceability as ``the potential to establish and use traces, representing the ability of an artifact or collection of artifacts to be tracked and understood throughout its life cycle''~\cite{gotel2012traceability}. In the context of software development, where evolving business needs and technological changes continuously drive new user requests, requirements are frequently modified. Traceability plays a crucial role in managing these changes by enabling the tracking of feature requests across the development lifecycle. It has been described as ``the ability to describe and follow the life of a requirement, in both a forward and backward direction.''~\cite{gotal1994analysis}

Traceability is commonly categorized into horizontal and vertical traceability. Horizontal traceability refers to links established among artifacts at the same abstraction level and development stage—for example, associations between related feature requests. In contrast, vertical traceability concerns the connections across different abstraction levels and development phases, such as linking a feature request to design documents, source code, or test cases.

Maintaining traceability is essential for ensuring consistency, supporting impact analysis, facilitating change management, and verifying whether stakeholder needs are fulfilled in the final product~\cite{de2008traceability}. In the following sections, we review existing work on feature request traceability from both horizontal and vertical perspectives.



\subsubsection{Horizontal Traceability of Feature Requests}
Horizontal traceability involves creating connections among software artifacts that exist at the same abstraction level or within a single development phase, particularly within RE~\cite{heck2014horizontal}. For feature requests, establishing horizontal traceability aids developers and stakeholders in gaining clear insights into project progress, efficiently managing the integration of new features, detecting requirement conflicts at an early stage, comprehending the rationale behind software requirements, handling interdependencies logically, and avoiding redundant or contradictory tasks, thereby reducing risks of software project failures.

Prior research on horizontal traceability for feature requests has focused on linking diverse artifacts, such as customer expectations with product requirements~\cite{matt2004speeding}, relationships between different feature requests~\cite{heck2014horizontal}, and connections among various issue reports~\cite{merten2016information}. Common relationship categories identified in the literature include duplicates, dependencies, conflicts, and references. Heck and Zaidman~\cite{heck2014horizontal}, for example, classified feature request relationships into dependencies (where resolution of one request precedes another), duplication, and reference. Similarly, Merten et al.~\cite{merten2016information} recognized relationships among issue reports as either duplicates (multiple reports describing identical software features) or generic (multiple reports referencing a common software feature).

Due to the considerable effort involved in manually creating these horizontal relationships, existing methods predominantly rely on automated techniques leveraging text similarity. For instance, Natt et al.~\cite{matt2004speeding} implemented a cosine similarity-based vector space model to effectively link customer expectations with corresponding product requirements, significantly enhancing efficiency. Heck and Zaidman~\cite{heck2014horizontal} employed TF-IDF and Latent Semantic Analysis (LSA) methods for text vectorization, demonstrating that TF-IDF notably outperformed LSA in identifying novel feature request connections. Furthermore, Merten et al.~\cite{merten2016information} evaluated various information retrieval algorithms, including TF-IDF, Latent Semantic Indexing (LSI), and variants of BM25, concluding that prioritizing feature request titles over comments substantially improves linking accuracy.

\subsubsection{Vertical Traceability of Feature Requests}
Vertical traceability focuses on creating connections between feature requests and artifacts across various phases or abstraction levels in software development~\cite{gotel2012traceability}. Implementing vertical traceability enhances developers' understanding of overall project status, facilitates rapid location of relevant artifacts across different development stages, and supports streamlined maintenance activities.

Licorish et al.~\cite{licorish2018linking} studied the associations between user requests, developer responses, and code modifications within the Android developer community. Their findings revealed strong associations between user requests and release notes, as well as between developer responses and corresponding code changes. However, direct links between user requests and source code modifications were notably weaker, suggesting indirect communication primarily through release notes.
To systematically support feature management, Seiler and Paech~\cite{seiler2017using} developed the Tagging Approach to Support Feature Management (TAFT), which leverages feature-specific descriptive tags to establish traceability. TAFT offers a lightweight yet structured mechanism beneficial for managing software features. However, the reliance on manual tag assignment introduces challenges related to consistency and representative accuracy.
Additionally, Palomba et al.~\cite{palomba2015user} introduced CRISTAL, a tool designed to link user comments extracted via AR-Miner~\cite{chen2014ar-miner} to relevant issue reports and commits occurring before subsequent software releases. CRISTAL employs asymmetric Dice similarity to establish these links, enabling clear tracking of user feedback implementation and facilitating responsive software evolution.

In summary, vertical traceability of feature requests plays a critical role in effectively assessing the impacts of changes to requirements, identifying affected components and documentation, evaluating potential risks associated with requirement modifications, and defining precise tasks required for successful implementation.

\subsection{Developer Assignment}
Once a feature request is approved, a critical step is assigning it to a suitable developer. Due to variation in developer expertise, proper assignment plays a vital role in reducing implementation failures and improving development efficiency~\cite{mao2015developer}. Most developer assignment techniques are based on the assumption that ``either the original author of the code or a developer with similar prior experience is the best candidate for the task.''

\textcolor{black}{Developer assignment is conceptually rooted in two long-standing research streams that address the same core challenge of matching a natural-language issue description to an appropriate expert: bug report assignment and ticket routing.}

\textcolor{black}{Bug report assignment originated from IR-based lexical matching between bug reports and developer artifacts~\cite{matter2009assigning, anvik2011reducing}. A seminal study by Anvik et al.~\cite{anvik2006should} first formulated bug report assignment as a supervised text classification problem, training models such as NB and SVM on historical project data. Each report was represented using textual and categorical features (e.g., summary, description, and component). The trained models predicted the most likely developers for new reports.
Over the years, techniques have gradually evolved toward data-driven and semantic reasoning, incorporating historical fixes, severity, and activity traces into ranking-based and deep neural models~\cite{xuan2014towards, tian2016learning, lee2017applying, mani2019deeptriage}.
Recent advances further leverage scheduling and dependency constraints, contrastive objectives, and large-scale industrial evaluations, transforming bug triage into a robust, deployable automation process~\cite{jahanshahi2022s, borg2024adopting, wang2024empirical}.}

\textcolor{black}{Ticket routing, emerging from enterprise service management, advanced from probabilistic and sequence-mining models for routing efficiency~\cite{shao2008efficient, shao2008easyticket} to graph and embedding-based expert representations that capture both technical competence and organizational relations~\cite{zhou2016resolution, xu2018expert, mandal2019automated}.
More recent systems leverage LLMs to jointly predict responsible teams and resolvers while exploiting organizational knowledge for retrieval-augmented and relation-aware routing~\cite{feng2022tadaa, zangari2023ticket, liu2025tickit}.}

\textcolor{black}{Together, these developments provide the conceptual and methodological groundwork for contemporary research on feature request assignment, where the focus shifts from fixing to implementing new functionality while preserving the same principle of expertise-oriented task routing.}

Feature request assignment approaches can generally be categorized into three groups:

\noindent  \textbf{\ding{46} Developer-Centric Approaches.}
These methods identify the most appropriate developer directly from the developer pool based on the textual similarity between the feature request and the developer's historical activity.
Canfora and Cerulo~\cite{canfora2006supporting} matched new feature requests with documentation generated from developers’ past commits and feature request descriptions. Developers with the most similar documentation are ranked and recommended. However, this approach depends heavily on the availability of rich historical data, which may not exist in younger or less active projects.

\noindent  \textbf{\ding{46} Artifact-Centric Approaches.}
These methods first retrieve source code artifacts (e.g., files, methods, classes) related to the feature request, and then identify developers associated with those artifacts.
Kagdi et al.~\cite{kagdi2009who, kagdi2012assigning} created a document index of code units using identifiers and comments, and employed LSI to match requests with code components. Developers were ranked based on their contribution distance, calculated using the Euclidean distance between commit vectors.
Linares-Vásquez et al.~\cite{linares2012triaging} improved accessibility by avoiding repository mining, instead linking requests to authors extracted from annotated source files using srcML~\cite{collard2013srcml}.
Hossen et al.~\cite{hossen2014amalgamating} introduced the iMacPro method, which ranks relevant code components by historical modification frequency and extracts both maintainers and original authors for recommendations. It demonstrated superior recall compared to earlier methods, such as iA~\cite{linares2012triaging} and iAcPro~\cite{hossen2014amalgamating}.
Zanjani et al.~\cite{zanjani2015using} proposed iHDev, which used TF-IDF and KNN to identify relevant code units, then used contribution history to rank developers. It outperformed xFinder~\cite{kagdi2012assigning} and iMacPro~\cite{hossen2014amalgamating} in both recall and Mean Reciprocal Rank (MRR).

\noindent
\textbf{\ding{46} Hybrid and Rule-Based Approaches.}
Combining repository mining with human-defined rules and external metadata offers a flexible yet robust solution.
Cavalcanti et al.~\cite{cavalcanti2016towards} presented a semi-automated rule-based expert system (RBES) combined with a machine learning classifier. The RBES incorporated organizational knowledge, while the machine learning model handled historical patterns. When both systems failed, manual assignment was triggered. This hybrid system outperformed pure machine learning-based baselines.
Yang and Sun et al.~\cite{yang2016recommending, sun2017enhancing} introduced the EDR\_SI method, which enriches developer profiles with personalized code history, collaboration networks, and commit logs. It uses Collaborative Topic Modeling (CTM) to associate developers with relevant files for future assignments, aiming to strike a balance between experienced and novice developers.

While many recommendation methods favor developers with extensive experience, this bias can result in task overload for senior developers and missed training opportunities for newcomers. Therefore, future approaches should better balance developer assignment by considering factors such as workload, expertise, familiarity with the project, task difficulty, and even informal skills.
Claytor and Servant~\cite{claytor2018understanding} suggested incorporating developers’ extracurricular skills to inform more balanced and effective task assignment, ultimately maximizing organizational efficiency.

\subsection{Code Localization}
Code localization refers to the process of identifying relevant code components that must be modified or extended to implement a feature request. Since feature requests often imply functional extensions or improvements to existing systems, they require tracing back to specific parts of the source code. A dominant strategy in this area is to calculate the textual or semantic similarity between feature request content and code artifacts, ranking the most similar components as localization targets.

\nff{One of the earliest and most influential studies in this field is by Marcus et al.~\cite{marcus2004information}, who introduced an IR-based approach using LSI for concept and feature location. The method represents source code entities as textual documents and computes their semantic similarity with feature request descriptions in a latent space, enabling the identification of related code even without exact keyword matches. This seminal work laid the foundation for later IR-based and hybrid localization techniques.}

\textcolor{black}{Recent surveys have shown that IR-based methods in software maintenance have progressively evolved from surface-level textual retrieval to semantically enriched and structure-aware reasoning~\cite{niu2025deep}.
Feature location, a closely related and historically rich research area, has evolved through several major phases over the past two decades. 
Early studies conceptualized feature location as an IR problem, where natural-language descriptions of software features were matched to relevant source code entities based on textual similarity~\cite{xue2012feature, hoseini2014software, binkley2015enabling}. 
Subsequent research moved beyond pure textual analysis to incorporate structural, historical, and dynamic information, resulting in hybrid approaches that combined static dependencies, execution traces, and change histories to improve localization accuracy~\cite{peng2013improving, chochlov2017historical, font2017achieving}. 
With the rise of deep learning, feature location began to adopt semantic representation learning techniques, replacing manually engineered IR features with distributed embeddings that better captured conceptual relationships between feature descriptions and source code~\cite{corley2015exploring, razzaq2020empirical, perez2020comparing}. 
More recently, the field has advanced toward intelligent and hybrid feature location, integrating semantic, structural, and behavioral signals as well as traceability and feature characteristics to achieve more robust and context-aware localization~\cite{echeverria2021comparison, razzaq2021effect, mukelabai2023featracer}. 
This long-standing evolution, from IR-based retrieval to semantically enriched and hybrid reasoning, has directly shaped modern code localization research.}

Building on this foundation, Palomba et al.~\cite{palomba2017recommending} proposed CHANGEADVISOR, which leverages user feedback for code localization. It analyzes comments with ARdoc~\cite{panichella2016app} to capture structure, semantics, and sentiment, then clusters feedback via LDA, LDA-GA, and HDP. Feedback clusters are linked to code classes using asymmetric Dice similarity; elements exceeding a threshold are ranked as likely implementation targets. The method achieves high accuracy and outperforms baselines in both precision and recall. \nff{Where2Change~\cite{zhang2019where2change} advanced CHANGEADVISOR by introducing Word2Vec to enrich semantics and applying a weight-selection cosine similarity to rank code classes, yielding higher Top-N accuracy and recall across 10 open-source apps.}
\nff{Furthermore, RISING~\cite{zhou2020user} leverages commit messages to bridge the vocabulary gap between reviews and source files, leading to better localization.}
\nff{Hu et al.~\cite{hu2022lighting} introduced a benchmark for user-review–based code localization: they curated gold-standard $\langle \text{review}, \text{source file} \rangle$ links for mobile apps and compared supervised with unsupervised approaches, with GraphCodeBERT consistently outperforming the baselines.
}
Ciurumelea et al.~\cite{ciurumelea2017analyzing} propose URR, which analyzes user reviews to localize change. Reviews are classified via Gradient Boosted Regression Trees into six high-level and twelve fine-grained categories. Code artifacts (files/classes/methods) are annotated and indexed, and Apache Lucene retrieves relevant files by matching processed reviews to the index, effectively mapping user concerns to implementation targets.
Kato and Goto~\cite{kato2017user} propose User-Generated Variables (UGV), a web-based IDE interaction model. UGV lets end users specify adjustable parameters without reading code; the system auto-links them to variable declarations, streamlining localization for non-technical stakeholders. This shifts interpretation to tooling, improving accessibility in feature-driven development. \nff{Recently, YourCoLo~\cite{chiyourcolo} is a user review-based code localization approach that models one-to-many review-to-code relationships and incorporates inter-code structural connections. It leverages project-level contextual signals and a customized ranking mechanism to more accurately identify relevant code elements linked to user feedback, outperforming SOTA baselines.}

\nff{Similar to code localization for feature requests, IR-based bug localization (IRBL)~\cite{zhou2012should, niu2025deep} aims to identify source code implicated by bug reports. Historical feature requests have been shown to provide valuable contextual signals for IRBL approaches~\cite{niu2023ablots, niu2024extensive, rath2018analyzing, yu2022towards}. Likewise, Xiao et al.~\cite{xiao2023reviewlocator} leveraged bug report information to enhance code localization for user reviews. Maarleveld et al.~\cite{maarleveld2025gotta} empirically examined the performance of IR-based file localization for issue reports (including both bug reports and feature requests), emphasizing the importance of models tailored to general issue localization and project-specific characteristics.}

\nff{Recently, NoCode-bench~\cite{deng2025nocode} benchmarks LLMs on natural-language-driven feature addition, bridging feature requests and code localization. The dataset contains 634 real tasks from 10 projects, pairing documentation updates with multi-file code changes. Nine state-of-the-art generative LLMs were evaluated, revealing their limited ability to map natural-language feature descriptions to the correct code regions. Besides, SWE-Bench~\cite{jimenez2023swe} and its extensions (including SWE-Bench Lite~\cite{jimenez2023swe}, SWE-Bench Multimodal~\cite{yang2024swe-multimodal}, and SWE-Bench Verified~\cite{chowdhury2024swebenchverified}) benchmark real-world issue resolution, encompassing both bug reports and feature requests. A series of LLM-based approaches~\cite{reddy2025swerank, xia2024agentless, chen2025locagent, yang2024swe} have been evaluated on these benchmarks to assess their capability in automating issue resolution tasks.}

\subsection{Code Recommendation}
Code recommendation methods help developers implement feature requests by suggesting relevant APIs or refactoring operations. Existing approaches mainly fall into two categories: learning from historical usage patterns or leveraging textual similarity.

\noindent
\textbf{\ding{46} API Recommendation.}
Thung et al.~\cite{thung2013automatic} matched feature requests with API method descriptions from historical software changes, achieving promising results across multiple libraries.
Xu et al.~\cite{xu2018muapi} proposed MULAPI, which integrates similarities among feature requests, source files, and API documentation to recommend APIs.
Sun et al.~\cite{sun2019enabling} enhanced MULAPI~\cite{xu2018muapi} by incorporating a CNN-based similarity model and simplifying parameter tuning, which significantly improved recommendation accuracy.
Li et al.~\cite{li2020analysis} highlighted API recommendation redundancy and advocated for evaluating API utility in conjunction with correctness.

\noindent
\textbf{\ding{46} Refactoring Recommendation.}
Niu et al.~\cite{niu2014traceability} leveraged feature-code traceability and code smells to recommend appropriate refactoring methods.
Nyamawe et al.~\cite{nyamawe2019automated, nyamawe2020feature} proposed a two-step classifier: first predicting whether refactoring is needed, then recommending specific techniques using multi-label SVM.
Nyamawe et al.~\cite{nyamawe2021identifying} also introduced a method to predict identifier renaming needs based on textual similarity between feature requests and source code.

\section{Publicly Available Datasets and Tools} \label{subsec:datasetandtool}
Bridging the gap between theoretical advancements and the practical application of feature request analysis requires tangible resources. The research community has responded by developing specific tools to implement proposed techniques and curating datasets derived from real-world sources, thereby facilitating empirical studies. This section catalogues these essential artifacts. 

\subsection{Tools for Feature Request  Analysis and Processing}
Translating research findings into practical applications and facilitating further empirical investigation necessitates the development of specialized tools. Numerous such tools, designed to assist with various stages of the feature request lifecycle, have been proposed in the literature. Table~\ref{tools_long} provides a consolidated overview of these tools, summarizing their intended application scenarios, the specific research themes they address, and their publication context. The existence of these tools, often targeting specific tasks like automated classification or prioritization, reflects the community's drive towards practical automation in handling feature requests.

\subsection{Publicly Available Datasets}
A significant impediment in many areas of RE research is the scarcity of publicly accessible datasets, often due to the proprietary nature of software requirements~\cite{yu2024makes}. However, the domain of feature requests presents a unique advantage: a substantial volume of data originates from open platforms, such as public repositories and app stores. This accessibility, often leveraged through web scraping techniques, has partially alleviated the data scarcity challenge that plagues other RE sub-fields. Consequently, various datasets have been curated and made publicly available by researchers. Table~\ref{tab:dataset_long} catalogs these resources, detailing their origins and the specific research tasks they are typically used for (e.g., training classification models, evaluating duplicate detection). 

{\small
\begin{longtable}{p{0.20\textwidth} p{0.28\textwidth} p{0.15\textwidth} p{0.18\textwidth} p{0.05\textwidth}}

\caption{Tools for Feature Requests Application.}
\label{tools_long} \\ 

\toprule
\textbf{Name} & \textbf{Description} & \textbf{Topic} & \textbf{Source} & \textbf{Year} \\
\midrule
\endfirsthead

\multicolumn{5}{c}{\tablename~\thetable\ -- \textit{Continued from previous page}} \\
\toprule
\textbf{Name} & \textbf{Description} & \textbf{Topic} & \textbf{Source} & \textbf{Year} \\
\midrule
\endhead

\midrule
\multicolumn{5}{r}{\textit{Continued on next page}} \\
\endfoot

\bottomrule
\endlastfoot

srcML~\cite{collard2013srcml} & Represents source code as XML for analysis and transformation. & Code Representation, Static Analysis, Transformation & \url{http://srcml.org} & 2013 \\
\addlinespace
CRISTAL~\cite{palomba2015user} & Establishes links between user feedback and submission records to analyze the impact on the development process. & Longitudinal Tracking & \url{http://www.cs.wm.edu/semeru/data/ICSME15-cristal} & 2015 \\
\addlinespace
URR~\cite{ciurumelea2017analyzing} & Analyzes and classifies user reviews to recommend code modifications. & Classification, Code Localization & \url{https://sites.google.com/site/changeadvisormobile/} & 2017 \\
\addlinespace
MARC2.0~\cite{nishant2018using} & Classifies application reviews and then generates summaries. & Classification, Summarization & \url{https://github.com/seelprojects/MARC-2.0} & 2018 \\
\addlinespace
piStar~\cite{pimentel2018pistar} & Provides a pluggable online platform for iStar modeling. & Goal Modeling & \url{https://github.com/jhcp/pistar} & 2018 \\
\addlinespace
Ticket Tagger~\cite{kallis2019ticket} & Tags GitHub issues as bug reports and feature requests. & Identification, Feature Extraction, Topic Clustering & \url{https://github.com/rafaelkallis/ticket-tagger} & 2019 \\
\addlinespace
ReFeed~\cite{kifetew2021automating} & Prioritizes software requirements by automatically analyzing user feedback properties. & Prioritization & \url{https://github.com/se-fbk/ReFeed} & 2021 \\
\addlinespace
SOLAR~\cite{gao2022listening} & Summarize helpful app reviews to facilitate release planning. & Identification, Summarization & \url{https://github.com/monsterLee599/SOLAR} & 2022 \\
\addlinespace
Izadi et al.~\cite{izadi2022predicting} & Predicts the objective and priority of software issue reports. & Classification, Prioritization, Issue Management & \url{https://github.com/MalihehIzadi/IssueReportsManagement} & 2022 \\
\addlinespace
VOTEBOT~\cite{wang2023pros} & Detects stance and summarizes opinions on feature requests. & Stance Detection, Summarization, Classification & \url{https://github.com/KeyL99/VoteBot} & 2023 \\
\addlinespace
Issue-Board~\cite{kegel2023automating} & Evolves domain models at runtime using textual feature requests. & Model Evolution, Language Engineering & \url{https://github.com/modicio} & 2023 \\
\addlinespace
Wei et al.~\cite{wei2024getting} & Get inspiration from the App Store and LLM-based approach for feature extraction. & Identification & \url{https://github.com/Jl-wei/feature-inspiration} & 2024 \\
\addlinespace
Khan et al.~\cite{khan2024mining} & Classifies user feedback into frequently occurring issue types. & Classification, Issue Detection & \url{https://github.com/nekdil566/issue-detection} & 2024 \\
\addlinespace
IARCT~\cite{biswas2024interpretable} & Classifies app reviews and explains the predictions using transformers. & Classification & \url{https://doi.org/10.5281/zenodo.11108088} & 2024 \\
\addlinespace
LLM-Cure~\cite{assi2024llm} & Suggests feature improvements by analyzing competitor user reviews. & Feature Enhancement, Recommendation, Competitor Analysis & \url{https://github.com/repl-pack/LLM-Cure} & 2024 \\
\addlinespace
Svensson et al.~\cite{svensson2024not} & Analyzes requirement prioritization criteria using Bayesian regression models. & Prioritization, Identification & \url{https://github.com/torkar/feature-selection-RBS} & 2024 \\
\addlinespace
T-FREX~\cite{motger2025leveraging} & Extracts mobile app features using encoder-only language models. & Feature Extraction, Classification, Identification & \url{https://github.com/gessi-chatbots/t-frex} & 2025 \\
\end{longtable}
}

{\small
\begin{longtable}{c p{0.49\textwidth} p{0.29\textwidth} c}

\caption{Open Sourced Datasets for Feature Requests.}
\label{tab:dataset_long} \\

\toprule
 \textbf{Ref.} & \textbf{Links} & \textbf{Description} & \textbf{Year}  \\
\midrule
\endfirsthead

\multicolumn{4}{c}{\tablename~\thetable\ -- \textit{Continued from previous page}} \\
\toprule
\textbf{Ref.} & \textbf{Links} & \textbf{Description} & \textbf{Year}  \\
\midrule
\endhead

\midrule
\multicolumn{4}{r}{\textit{Continued on next page}} \\
\endfoot

\bottomrule
\endlastfoot

\cite{wang2023pros}  & \url{https://github.com/KeyL99/VoteBot/} & \multirow{3}{=}{Identify feature requests from user review.} & 2023 \\ 
\cite{biswas2024interpretable}  & \url{https://doi.org/10.5281/zenodo.11108088} & & 2024 \\ 
\cite{khan2024mining}  & \url{https://doi.org/10.5281/zenodo.11256608} & & 2024 \\ 
\cline{1-4}

\cite{thorsten2016software}  & \url{https://zenodo.org/record/56907#.X-AES8gzabg} & \multirow{3}{=}{Identify feature requests from issue reports.} & 2016 \\
\cite{tao2019labelling}  & \url{https://github.com/heulaoyoutiao/bugtag2} & & 2019 \\ 
\cite{izadi2022predicting}  & \url{https://zenodo.org/record/4925855} & & 2022 \\
\cline{1-4}

\cite{bakar2016extracting}  & \url{https://www.dropbox.com/sh/mzd6yu1n8nckaod/AABi4wnmf8VR_90dc6C706cxa?dl=0} & \multirow{7}{=}{Feature extraction.} & 2016 \\
\cite{faiz2018the}  & \url{https://alt.qcri.org/semeval2014/task4/index.php?id=data-and-tools} & & 2018 \\ 
\cite{motger2025leveraging}  & \url{https://github.com/gessi-chatbots/t-frex} & & 2025 \\

\cite{wei2024getting} & \url{https://huggingface.co/datasets/Jl-wei/gp-app-description} & & 2024 \\ \hline

\cite{simone2019listening}  & \url{https://dibt.unimol.it/reports/clap/} & \multirow{3}{=}{Prioritization.} & 2017 \\
\cite{kifetew2021automating} & \url{https://github.com/se-fbk/ReFeed} & & 2021 \\
\cite{svensson2024not} & \url{https://github.com/torkar/feature-selection-RBS} & & 2024 \\ 
\cline{1-4}
\cite{sheikhaei2023study} & \url{https://doi.org/10.6084/m9.figshare.19382156} & Classification. & 2023 \\ \cline{1-4} 
\cite{gao2022listening} & \url{https://github.com/monsterLee599/SOLAR} & Summarization. & 2022 \\ \cline{1-4} 

\cite{palomba2015user} & \url{http://www.cs.wm.edu/semeru/data/ICSME15-cristal} & Traceability. & 2015 \\ \cline{1-4}
\cite{nyamawe2020feature} & \url{https://github.com/nyamawe/FR-Refactor} & Refactoring recommendation. & 2020 \\ \cline{1-4}
\cite{palomba2017recommending} & \url{https://sites.google.com/site/changeadvisormobile/} & Feature location. & 2017 \\ \cline{1-4}
\cite{niu2022towards} & \url{https://zenodo.org/record/6544368} & Approval prediction. & 2022 \\ \cline{1-4}
\cite{hu2022lighting} & \url{https://github.com/lcynju/review2code} & Feature location. & 2022 \\ 

\end{longtable}
}
\section{Challenges and Opportunities} \label{sec:challenges}
\subsection{Summarization}
This paper presents a systematic survey focused on the analysis and processing of user feature requests, acknowledging their critical role as a direct source of user needs for contemporary software evolution. Employing a defined research methodology that included systematic literature searching, rigorous screening protocols, and snowballing techniques (detailed in Section~\ref{methodology}), we identified and thoroughly analyzed \papers~relevant primary studies published up to August 2025.

Before delving into the core findings, the study established essential context by defining software features, feature requests, and related concepts, underscoring their value and distinguishing them from traditional, formally elicited software requirements. We also positioned this review in relation to existing surveys, clarifying its unique contribution in providing an end-to-end RE perspective on the feature request lifecycle (Section~\ref{relatedwork}).

The main body of our findings (presented from Section~\ref{sec:demographics} onwards) begins by outlining the demographics of the research landscape. This revealed a consistent growth in publications dedicated to feature requests since 2010, with significant activity concentrated in reputable software engineering and RE venues, confirming the topic's increasing importance within the community. We systematically cataloged the diverse origins of feature requests, confirming that while dedicated platforms like issue trackers are used, a vast amount originates from less structured channels such as app reviews, social media, and developer communications, with app reviews and issue reports being the most frequently studied sources (Section~\ref{sec:elicitation}).

Addressing RQ3, the core analysis examined how these requests are analyzed and processed within a RE framework. We reviewed various automated methods designed to identify feature request descriptions from often noisy and informal user feedback, spanning unsupervised, supervised, and semi-supervised paradigms. Then we discussed techniques for software feature extraction, often relying on lexical approaches like POS tagging. It also included methods for structural analysis aimed at understanding content, such as user intent recognition, classification into diverse functional or non-functional categories, and automated summarization. Furthermore, the critical task of prioritization was examined, contrasting traditional RE approaches with methods adapted specifically for the high volume and nature of user feedback. Moreover, we examined approaches for quality review, primarily focusing on quality assessment related to duplication, redundancy, and ambiguity. We also explored traceability methods, encompassing both horizontal links between related requests and vertical links to artifacts such as source code and developers. Finally, various facets of change management were reviewed, including predicting request approval, recommending suitable developers, localizing relevant code sections, and suggesting APIs or refactorings.

A cross-cutting theme throughout the analysis and management stages was the pervasive application of NLP and artificial intelligence (AI) techniques, driven by the necessity to automate the handling of large volumes of unstructured text. Recognizing the importance of facilitating further research and reproducibility, this study also compiled publicly available resources, specifically cataloging tools developed for feature request processing and relevant datasets derived from open sources, classified by the research topics they support. 


\subsection{Challenges and Opportunities}
While this systematic literature review provides a snapshot of the current landscape, the dynamic nature of software development and user feedback necessitates continuous investigation. We intend to actively monitor the evolving research in feature request analysis and processing to reflect new advancements and shifting trends.
Beyond ongoing monitoring, several specific avenues represent critical next steps for the research community. 

\noindent \textbf{Quality Assurance of Feature Requests.} 
One critical yet underexplored dimension is the quality assurance of feature requests. As our study reveals, many feature requests are written in natural language and often suffer from common quality issues such as ambiguity, incompleteness, redundancy, or inconsistency~\cite{shi2016new, heck2013analysis}. These so-called quality smells hinder effective downstream tasks such as prioritization, clustering, and specification. Future research should prioritize the development of tools and techniques to detect and classify quality smells in user-submitted feature requests automatically. Building on existing work in requirements quality assessment and NLP, such tools could flag problematic requests early in the processing pipeline.

Beyond detection, there is an urgent need for approaches to guide the improvement or refinement of low-quality feature requests. By systematically addressing the quality of feature requests, the community can unlock more reliable and effective methods for leveraging user feedback in software evolution.

\noindent \textbf{Comparative Analysis of Feature Requests and Bug Reports.}
Firstly, this review has focused exclusively on feature requests. However, user feedback often intertwines requests for new functionality with reports of existing problems~\cite{dkabrowski2022analysing}. A crucial future direction is to integrate the study of bug reports alongside feature requests. Adopting a RE lens, comparative analyses are needed to explore the similarities and differences in effectively analyzing, prioritizing, and managing these two distinct yet interconnected types of feedback. Understanding where techniques converge and where they must diverge could lead to more holistic and efficient feedback management frameworks and tooling.

\noindent \textbf{Specification and Validation of Feature Requests.}
Furthermore, a significant and largely unaddressed gap identified in our study pertains to the specification and validation of feature requests. Currently, research heavily emphasizes identification and analysis, but largely overlooks the critical subsequent steps needed to bridge the gap between raw user input and actionable, verifiable requirements. Future research must urgently tackle the challenge of transforming collections of informal, often ambiguous, and fragmented user requests into more structured, consistent, and usable requirement representations. This could involve exploring techniques for semi-automatically generating preliminary user stories, use case scenarios, behavioral models, or even identifying constraints and non-functional aspects implicitly mentioned in feedback. Equally vital, and even less explored, is the validation of these derived specifications~\cite{mallya2026rita}. Methodologies and practical techniques are needed to ensure that the interpretations and specifications derived from user feedback accurately reflect the original user needs and intent before substantial development effort is invested. Addressing this specification and validation gap is fundamental to truly leveraging feature requests for building software that meets genuine user expectations. We plan to conduct in-depth investigations into these specific areas in our future work.

\noindent \textbf{Leveraging LLMs for Feature Request Intelligence.}
With the rapid advancement of LLMs, their application in software engineering tasks has gained increasing traction~\cite{yu2024makes}. However, our review finds that feature request research has yet to leverage the potential of LLMs fully. These models offer promising capabilities for a wide range of tasks, including classifying user intent, mapping feature requests to relevant source code locations, summarizing feedback, and recommending suitable developers or implementation strategies. Future work should explore how LLMs can be fine-tuned or prompted for domain-specific use in feature request processing, and investigate their effectiveness compared to traditional machine learning techniques. Moreover, new evaluation criteria and human-in-the-loop workflows may be necessary to ensure reliability, interpretability, and alignment with user expectations.

\noindent \textbf{Lack of Benchmarks for LLM-Driven Feature Request Tasks.}
Despite growing interest in applying LLMs to software maintenance and requirements tasks, there is a notable absence of standardized benchmarks specifically targeting LLM-based feature request processing~\cite{hu2025assessing}. This limits reproducibility, fair comparison across methods, and progress tracking over time. To address this, the research community should work toward building and releasing well-curated, multi-task benchmark datasets that reflect real-world feature request scenarios. These benchmarks should include annotations for classification, prioritization, code mapping, and developer assignment, and be accompanied by baseline LLM implementations to support further development and evaluation.

\noindent \textbf{Multilingual and Cross-Cultural Feature Request Processing.}
Current research on feature requests is heavily skewed toward English-language feedback~\cite{bhuiyan2026nonenglish}, predominantly from Western platforms and user bases. However, in today’s global software market, products often receive feedback in multiple languages and cultural contexts. This introduces new challenges in understanding linguistic nuances, implicit user expectations, and regional preferences. Future work should explore multilingual models and culturally adaptive techniques to support accurate identification and interpretation of feature requests across diverse user groups. This direction also calls for the development of multilingual corpora and evaluation tools.

\noindent \textbf{Incorporating Temporal Dynamics in Feature Request Analysis.}
Feature requests evolve over time: some lose relevance, while others gain traction with emerging trends or shifts in business priorities. Yet, few existing studies explicitly model the temporal aspect of feature request lifecycles. Incorporating temporal information, such as request aging, frequency of similar requests over time, or alignment with product release schedules, can enhance prioritization and planning. Future research could explore temporal modeling techniques (e.g., time-aware neural networks, survival analysis) to support more dynamic and responsive feature request management.

\noindent \textbf{Bridging User and Developer Perspectives.}
Feature requests often originate from non-technical users but are consumed and implemented by developers, creating a semantic and expectation gap. Bridging this gap remains an underexplored area. Future work could investigate how to translate user-facing language into developer-understandable representations, integrate user sentiment and urgency into development planning, and develop tools that facilitate two-way communication between stakeholders. Enhancing alignment between user needs and developer actions could significantly improve the effectiveness and user satisfaction of feature implementation~\cite{bano2017user}.

\noindent \textbf{Autonomous Agents for End-to-End Feature Request Processing.}
Another emerging and promising direction for future research is the development of autonomous agents capable of handling feature requests throughout the entire lifecycle, from initial user interaction and intent interpretation to analysis, code localization, and implementation. With advancements in multi-agent collaboration frameworks and AI planning~\cite{zhang2025survey, tran2025multi}, it is increasingly feasible to envision intelligent agents that can process user feedback, generate actionable development tasks, and even coordinate with other agents for implementation and testing. Such agent-based systems could significantly reduce the human effort required for managing feature requests, especially in large-scale and user-driven software systems. Exploring the design, capabilities, and limitations of such autonomous feature request agents offers a valuable avenue for advancing intelligent software engineering.

In summary, while significant progress has been made in feature request research, numerous challenges remain that hinder its full potential in modern software engineering. From the integration of bug reports and the specification-validation gap, to the untapped potential of LLMs, lack of benchmarks, multilingual considerations, temporal dynamics, and the promising vision of agent-based automation—each represents both a pressing challenge and a valuable opportunity. Addressing these issues can pave the way for more intelligent, scalable, and user-aligned approaches to feature request processing, ultimately contributing to the development of software systems that more effectively meet the real needs of users.
\section{Threats to Validity}\label{threats}
This section outlines the threats to validity, categorized into four dimensions: construct validity, internal validity, external validity, and conclusion validity.

\noindent
\textbf{Construct Validity.}
The primary threat to construct validity stems from the selection and inclusion of studies. While we conducted comprehensive searches across major software engineering digital libraries using carefully formulated keywords, there remains a risk that some relevant studies were missed due to limitations in the search terms or database indexing. To mitigate this risk and broaden our study coverage, we also employed both forward and backward snowballing. Another potential threat lies in the application of inclusion and exclusion criteria, which may introduce subjective bias. To address this, all candidate studies were independently reviewed by multiple authors based on predefined selection criteria. Discrepancies were resolved through discussion, and a third author was consulted when necessary to reach consensus.

\noindent
\textbf{Internal Validity.}
Researcher bias may influence judgment during the study selection and data extraction process. To mitigate this, the inclusion and exclusion criteria were rigorously discussed and agreed upon prior to their application. A dual-review process was implemented to cross-validate decisions. Any disagreements were resolved through author discussions, with a third author involved when consensus could not be reached.

\noindent
\textbf{External Validity.}
The scope of the included studies constrains the generalizability of our findings. This review specifically targets research involving feature requests within the context of software engineering practices. Broader studies on user feedback or those situated outside the software engineering domain were excluded. Consequently, the applicability of our results may be limited when extended to other forms of user-centered RE.

\noindent
\textbf{Conclusion Validity.}
Our conclusions are grounded in the analysis of the selected primary studies. However, differences in individual interpretations during the selection and coding phases could lead to variations in results if other researchers were to repeat the process. To minimize this risk, we conducted regular discussions among the authors to ensure a shared understanding and reduce interpretation bias.

\section{Conclusion}\label{conclusion}
User feature requests, sourced directly from end-users, are increasingly vital for software evolution but pose unique challenges due to their volume, informality, and divergence from traditional requirements. This paper presents a comprehensive survey of \papers~studies, synthesizing the current research landscape on analyzing and processing these requests throughout the software lifecycle. We examine techniques for twelve research topics on feature requests, organizing findings within a RE framework. The survey concludes by highlighting significant research gaps and future directions.

\begin{acks}
This work was supported by the National Natural Science Foundation of China (61802167).
\end{acks}

\bibliographystyle{ACM-Reference-Format}

\appendix

\end{document}